\documentclass{article}
\usepackage{graphicx} 
\usepackage{amsmath} 
\usepackage{amsfonts} 
\usepackage{subfigure}

\begin{document}

\title{Discrete Choices under Social Influence: \\Generic Properties}
\author{Mirta B. Gordon {\small (1)}, Jean-Pierre Nadal {\small (2,3)}, \\Denis Phan {\small (4,5)}, Viktoriya Semeshenko {\small (1)} }
\vspace{2cm}
\date{{\small
(1) Laboratoire Techniques de l'Ing\'enierie M\'edicale et de la Complexit\'e \\ 
(TIMC-IMAG, UMR 5525 CNRS-UJF), Universit\'e Joseph Fourier, Grenoble \\
(2) Centre d'Analyse et Math\'ematique Sociales (CAMS, UMR 8557 CNRS-EHESS), \\
Ecole des Hautes Etudes en Sciences Sociales, Paris \\
(3) Laboratoire de Physique Statistique (LPS, UMR 8550 CNRS-ENS-Paris 6-Paris 7), \\ 
Ecole Normale Sup\'erieure, Paris\\
(4) Centre de Recherche en Economie et Management \\ 
(CREM, UMR 6211 CNRS-Universit\'e de Caen-Universit\'e de Rennes 1), Universit\'e de Rennes 1 \\
(5) Groupe d'Etude des M\'ethodes de l'Analyse Sociologique \\ (GEMAS, UMR 8598 CNRS-Universit\'e Paris Sorbonne - Paris IV), Paris \\
$\;$ \\$\;$} 
March 8, 2007 \\}

\maketitle


\begin{abstract}
We consider a model of socially interacting individuals that make a binary choice in a context of positive additive endogenous externalities. It
encompasses as particular cases several models from the sociology and economics literature.
We extend previous results to the case of a general distribution of  idiosyncratic preferences, 
called here Idiosyncratic Willingnesses to Pay (IWP).
 
Positive additive externalities yield a family of inverse 
demand curves that include the classical downward sloping ones but also new ones with non constant convexity.
When $j$, the ratio of the social influence strength to the standard deviation of the IWP distribution,
is small enough, the inverse demand is a classical monotonic (decreasing) function of the adoption rate.
Even if the IWP distribution is mono-modal, there is 
a critical value of $j$ above which the inverse demand is non monotonic, 
decreasing for small and high adoption rates, but
increasing within some intermediate range. 
Depending on the price there are thus either one or two equilibria. 

Beyond this first result, we exhibit the {\em generic} properties of the boundaries limiting the regions 
where the system presents different types of equilibria (unique or multiple).
These properties are shown to depend {\em only} on qualitative features of the 
IWP distribution: modality (number of maxima), smoothness
and type of support (compact or infinite).
The main results are summarized as {\em phase diagrams} in the space of the model parameters, 
on which the regions of multiple equilibria are precisely delimited.

\end{abstract}

\newpage

\tableofcontents

\newpage

\section{Introduction}
\label{sec:intro}

\subsection{Modeling social influences with heterogeneous agents}
\label{sec:general}
There are many circumstances in social and economic contexts where, faced with different alternatives, the best choice for an  individual depends on the choices of other individuals in the population. The decision of leaving a neighborhood \cite{Schelling}, to attend a seminar \cite{Schelling} or a crowded bar \cite{Arthur94,Arthur99}, to participate to collective actions such as strikes and riots \cite{Granovetter78}, are particular examples taken from social sciences. 
It has been suggested that social interactions may explain the school dropout \cite{Crane91}, the persistence in the educational level within some neighborhoods \cite{Durlauf96} and the related consequences in the stratification of investment in human capital and economic segregation \cite{Ben96}, the large dispersion in urban crime through cities with similar characteristics \cite{GlaeSaceSche}, the emergence of social norms \cite{Ostrom00}, the labor market behavior and related unemployment patterns \cite{Top01,Con02}, the housing demand \cite{Ioa03}, the existence of poverty traps \cite{Durlauf06}, the smoking behavior \cite{Kra06a,Kra06b,Soe06}, etc.

Similarly, there is a growing economic literature that recognizes the influence on consumers of the social world they live in. In market situations like the subscription to a telephone network \cite{Art73, Roh74, vRab74,Cur87} or the choice of a computer operating system \cite{KAT94}, the willingness to pay generally depends not only on the individual preferences but also on the choice made by others \cite{ShapVar99,Roh01}. 
If the externality is positive the utility of the most popular choice increases even for individuals who otherwise would never make this choice. 
In other words, the {\em conformity effect} may dominate the {\em heterogeneity of preferences}, as pointed out by Bernheim \cite{Ber94}. General aspects of these issues have been discussed in the literature \cite{Bec00,Man00}. Particular insightful papers are Becker's note \cite{Becker91} about restaurants pricing, and the qualitative analysis by Granoveter and Soong \cite{GranovetterSoong86} of the consequences of interpersonal influences (``bandwagon effects"\cite{LIE50,Roh01}) on the consumers demand and on the supply prices.

In the present paper we consider the general properties of a model of socially interacting heterogeneous individuals that make a binary choice in a context of positive endogenous externalities. The model encompasses, as particular cases, most of the above mentioned models presented in the sociology and economics literature. 
In a forthcoming paper \cite{mono} we explore the consequences of the externalities on the economy, taking as an example the simplest market, i.e. that of a monopolist pricing a single good.

In social sciences, the question of discrete (typically binary) choices with heterogeneous agents and positive externalities has been first addressed in the 70's by Schelling \cite{Schelling73, Schelling}, who borrowed from Physics the concept of {\em critical mass}: in a repeated-decisions setting, depending on whether this critical mass is or not reached, the system may end up at very different equilibria. 
Granovetter further develops Schelling's model, applying it to particular problems such as joining or not a riot \cite{Granovetter78}, voting, etc \cite{GranovetterSoong86}. 
The same topic is reconsidered within a statistical physics point of view in the early 80's by Galam {\em et al} \cite{GaGeSh}. The notion of critical mass is then related to the Physics concept of {\em phase transition} at a critical point, in the neighborhood of which the system may be extremely susceptible:
by tipping effects, small microscopic changes can lead to drastic changes
at the macroscopic level. Similar tools have been applied in 1980 by Kindermann and Snell \cite{KindSnell} to the study of social networks. These authors introduced into the sociology and economics literature the equivalence between statistical physics approaches ---that use the Boltzmann-Gibbs distribution--- and Markov Random Fields. 
Another physically-inspired approach
for modeling social phenomena such as opinion diffusion has been developed by Weidlich and Haag \cite{WeidlichHaag83,Wei00} in 1983, through a master equation and the Fokker-Planck approximation. 
Later, these physically inspired models of opinion contagion have been exploited in economics by Orl\'ean \cite{Orlean90,Orlean95} for the analysis of mimetic behaviors in the context of financial markets.  
There is now a large and growing literature on opinion and innovation diffusion (see e.g. \cite{Valente,Schweitzer01,Deffuant01,WattsCascades,SocialPerco}) 
closely related to the general discrete choice model considered in this paper. 
Since the beginning of the 90's the general framework of social interactions in
non-market contexts is reconsidered in a
 Beckerian way \cite{Bec74,Bec00}, in particular by Glaeser {\em et al}
\cite{GKSS92,GlaeSaceSche,GlaeserScheinkman}.

The first application of statistical mechanics approaches in economics may be traced back to the pioneering work of F\"{o}lmer \cite{Fol74}.  Introducing an economic interpretation of the Ising model of ferromagnetism at finite temperature, he shows that strong externalities may hinder the stabilization of an economy. These models introduce Markov random fields (equivalently Boltzmann-Gibbs distributions) to model uncertainty in the decision making process, allowing for the definition of a general equilibrium concept. According to F\"{o}lmer, Hildebrand's \cite{Hildebrand83} justification 
of the representative agent approach breaks down when agents' decisions are correlated due to their social interactions (for a discussion, see also \cite{Kirman92}). 

A renewal of interest for models of binary decisions with externalities arose in economics in the 90's. On one side, Durlauf and collaborators \cite{Durlauf91,Durlauf93,Durlauf94,Durlauf96} and Kirman and Weisbuch \cite{WeisbuchKirman} among others, consider agents that choose an action according to a Boltzmann-Gibbs distribution, that is a {\em logistic} choice function, reflecting some random aspects in the agent's utility. In this context Brock \cite{Brock93} and Blume \cite{Blume93,Blume95} explicit the links between Game Theory and Statistical Mechanics, while Kirman and coworkers \cite{NaWeChKi} show that the logistic choice function may be seen as resulting from an exploration-exploitation compromise. These and other recent papers \cite{ArDuLa,Durlauf97,Durlauf99,BrockDurlauf01a,BrockDurlauf01b,GlaeserScheinkman,WeiSta,EhMaVe04,GoNaPhVa05,MichardBouchaud,BorghesiBouchaud} analyze with statistical physics tools the consequences of positive social (market and non-market) interactions in the aggregate behavior of large populations (for a short introduction to statistical physics approaches see \cite{Gordon04} and for their application to economics see \cite{PhaGorNad04}; see also \cite{Ball} for a survey). 
Most of the above mentioned authors
restrict the analysis of the model to the case where all the individuals have the same idiosyncratic preference. Heterogeneity in the population is introduced through the probabilistic decision-making process ({\em random utility model} \cite{Luce59,Manski77}), like in F\"{o}lmer's work \cite{Fol74}. Then, the actual equilibrium reached by the system depends on the fixed points of the decision dynamics, generally a myopic best reply. An interesting characteristic of these models is that they present multiple equilibria for some range of the parameters. Becker \cite{Becker91} pointed out important consequences of these multiple equilibria, induced by externalities, on the economy: he suggests that they could be the reason of seemingly suboptimal pricing in situations of persistent excess demand. 

In this paper we consider intrinsically heterogeneous agents with fixed utilities, like in McFadden's approach to Quantal Choice models \cite{McFadden74,McFadden76}. Each individual has an {\em Idiosyncratic Willingness to Pay} (IWP) that remains fixed on time. 
We mainly (but not exclusively) decline the model within a market context, in which the binary choice corresponds to buying or not a given good at a posted price.
This general setting allows us to generalize Becker's qualitative analysis \cite{Becker91} of the optimal pricing problem. Putting the price to zero allows us to recover the social sciences models. We assume that these IWP are random variables that are distributed among the population according to a given probability density function (pdf).

We determine the possible equilibria of the system without assuming any precise decision-making dynamics. We show that the model's properties depend on the strength of the externality and on qualitative properties of the IWP pdf, like its modality class (the number of maxima), its smoothness properties and the kind of support. We display the main results on a plane whose axes are the parameters of the model, namely, the average IWP and the strength of the social component, both measured in units of the standard deviation of the IWP distribution. Particular cases of our model have been published elsewhere \cite{NaPhGoVa05,GoNaPhVa05}. This paper extends those result to the case of a general IWP distribution.
The particularly important case of a uni-modal pdf (with a single maximum) is thoroughly studied, but we also discuss the consequences of multi-modality.
Our results are summarized on {\em phase diagrams}, that is,
we plot in the parameters space the boundaries of the regions where different types of solutions exist.

Before entering into the details of our approach (section \ref{sec:model}) we discuss the analogies and differences between our model and other models of social interactions (section \ref{sec:recuit}) and we summarize our main results (section \ref{sec:mainresults}).

\subsection{More on related models}
\label{sec:recuit}
In this section we briefly discuss the relationship between the model to be considered here and other models studied in the literature. 
Let us first consider models of discrete choices in the absence of externalities.
According to the typology proposed by Anderson {\it et al.} \cite{AnDeTh}, within the general framework of {\em Random Utility Models} (RUM)\cite{Luce59,Manski77} with  {\it additive} stochastic utilities, there are two distinct approaches to individual choices: a ``psychological" one and an ``economic" one. In the psychological perspective (Thurstone \cite{Thurstone}, Luce \cite{Luce59}) the randomness is a time-dependent i.i.d. random variable: the random components of the idiosyncratic preferences are assumed to be independently drawn afresh by each individual from a given pdf, each time the choice has to be made. They are interpreted as individual temporary changes, or mistakes in the estimated utilities. In the simplest case ---actually, the only one treated in the social and economic literature--- the agents IWPs have identical deterministic parts and only differ by this random time-varying term which is systematically assumed to be drawn from a logistic pdf. 
In practice many approaches like in \cite{BrockDurlauf01a} consider the choice rule
as deriving from a random utility model \cite{Luce59}. As shown by McFadden \cite{McFadden74}, in this context
the logistic form
is obtained if the random terms in the underlying Thurstone's discriminant process are i.i.d. Weibull random variables, i.e. have a double exponential (extreme value, type I) distribution (see also \cite{AnDeTh}).

In the presence of strategic complementarities (\cite{BuGeKl,Cooper99}), the resulting model is well known in statistical physics: it corresponds to the {\em standard Ising model}, i.e. with ferromagnetic interactions and {\em annealed} disorder, that is, at finite temperature $T$. The latter is the inverse of the standard logistic parameter $\beta \equiv 1/T$ and is thus proportional to the standard deviation of the IWP distribution. The ferromagnetic interaction constant $J$ corresponds to the strength of the social externality. Introduced by F\"{o}lmer \cite{Fol74} in the economics context, this standard Ising model has recently been reconsidered in the social and economic literature mainly by Durlauf and coworkers \cite{Durlauf97,Durlauf99,BrockDurlauf01a,BrockDurlauf01b} and by Kirman and Weisbuch \cite{WeisbuchKirman,NaWeChKi}. The corresponding equilibria are reminiscent of the Quantal Response Equilibria \cite{McKelveyPalfrey95} used in the context of experimental economics and behavioral game theory. These are equilibria ``on the average", in the statistical sense (as in Physics): they do not correspond to the strict maximization of the utilities (that are random variables) but to that of their estimated or expected values. In the generally considered infinite population limit 
(where the variance of the expected values vanishes) 
the expected utilities  are systematically smaller that the maximal ones.

The standard Ising model is quite well understood \cite{Stanley}. Ising himself \cite{Ising24} gave an analytic description of its properties in the simplest case of a linear chain where each agent interacts only with his right and left nearest neighbors. There is also an analytical description of the stationary states of the model, due to Onsager \cite{Onsager}, in the case where the agents are situated on the vertices of a 2-dimensional square lattice, each having four neighbors. However, no analytic results exist for arbitrary neighborhoods except for the specific case of a global neighborhood, known as the {\it mean field} approximation in Physics. Accordingly, considering global neighborhoods, Brock and Durlauf \cite{BrockDurlauf01a} analyze the properties of the expected demand in the case of rational expectations under the assumption of a logistic distribution of such expectations (assuming thus double exponential random utilities). They find, in agreement with standard results in statistical mechanics \cite{Stanley}, that there exist either one, two or three solutions for the demand function, depending on the relative magnitudes of the idiosyncratic uniform social term, the variance of the stochastic term and the strength of the social effects. 

In the following we adopt instead McFadden's \cite{McFadden74} economic approach (see also 
\cite{Manski77,AnDeTh}): we assume that each agent has a willingness to pay {\it invariable} in time, that is different from one agent to the other. In statistical physics this heterogeneity is called {\em quenched} disorder. The particular model we study is analogous to the ferromagnetic {\em Random Field Ising Model} (RFIM) at zero temperature (corresponding to the fact that the agents make deterministic choices). Thus, our modeling approach assumes the so called ``risky'' situation: an external observer (e.g. a seller) does not have access to the individual preferences, but may know their probability distribution. 
According to McFadden, ``Thurstone's construction is appealing to an economist because the assumption that a single subject will draw independent utility functions in repeated choice settings and then proceed to maximize them is formally equivalent to a model in which the experimenter draws individuals randomly from a population with differing, but fixed, utility functions, and offers each a single choice; the latter model is consistent with the classical postulates of economic rationality'' (\cite{McFadden76}, p 365). However, in the presence of social interactions this statement is in general incorrect. In a repeated choice setting, individual utilities evolve in time according to the others' decisions. In fact, the equilibria of systems reached through a dynamics that corresponds to an iterated game where agents make myopic choices at each time-step is one of the main concerns of statistical physics. It is known that the equilibria of systems with annealed or quenched disorder are of very different nature. The time average on a single agent and the population average at a given time do not necessarily coincide. 

In contrast to the standard Ising model at finite temperature (annealed disorder), the properties of the RFIM with externalities, both at zero and at finite temperatures, are far from being fully understood. The properties of quenched disordered systems have been and are still the subject of numerous studies. Since the first studies of the RFIM, which date back to Aharony and Galam \cite{GalAha80,Galam82}, a number of important results have been published in the physics literature on this model (see e. g. \cite{Sethna93}). Several variants of the RFIM have already been used in the context of socio-economic modeling, both by physicists and economists \cite{GaGeSh,Orlean95,Bouchaud,WeiSta}.  

The quenched-utilities model (RFIM at zero temperature) and the annealed-disorder model (standard mean field Ising model at finite temperature) have the same aggregate behavior (i.e. demand function for the market case) and equilibria under the following conditions - but essentially {\em only} under such conditions:
\begin{enumerate}
\item the choice function with annealed utilities is identical to the cumulative distribution of the quenched IWPs; 
\item in the annealed case, equilibrium is reached through repeated best reply choices, where the expected demand is myopically estimated;
\item the population size is infinite, guaranteeing that the variance of the demand vanishes in both models.
\end{enumerate}

However, the economic interpretation of these equilibria are very different: in the case of quenched utilities these are standard Nash equilibria, while in the case of annealed utilities these are similar (although not identical) to Quantal Response equilibria.

\subsection{Main specific results}
\label{sec:mainresults}
In the present paper we determine the {\em equilibrium properties} in the case of
a {\em global neighborhood} with time invariant (quenched) random utilities, in the limit of an infinite number of agents. Since our paper focuses on equilibrium (static) properties, the social influence depends on the {\em actual} choices of the neighbors, in contrast with \cite{BrockDurlauf01a}, where the social influence in the surplus function depends on the agent's {\em expected} demand. 

Previous studies using annealed or quenched utilities consider specific probability distributions, mostly a logistic or a Gaussian \cite{Thurstone}. Some papers have determined conditions on the choice function for having multi-equilibria \cite{NaWeChKi,GlaeserScheinkman}. From the Physics literature we expect that specific properties near a {\em critical point} (a bifurcation point, see Section \ref{sec:boundary} below) are independent of the details of the model: this is used in \cite{Sethna93} for describing the hysteresis effects in a family of (physical) systems at such a critical point, and exploited in \cite{MichardBouchaud} for the analysis of empirical socio-economic data in cases where the actual pdfs are not known. However, the full description of the {\em phase diagram} for an arbitrary pdf has not been done yet. Here we present this detailed analysis for a typical probability distribution of the IWP. We show how uniqueness or multiplicity of equilibria, related to convexity properties of the inverse demand functions, result from modality and smoothness of the pdf, as well as from the strength of the externality.

We show that for small enough strengths of the social influence (the case of moderate social influence in \cite{GlaeserScheinkman}), the demand has a classical shape, that is, with a continuous decreasing adoption 
rate for increasing prices. However, if the ratio between the social influence strength and the standard deviation of the IWP distribution exceeds a critical value, the inverse demand function exhibits a non-classical, non-monotonic, behavior.
As a result, depending on the price, there are either one or two stable equilibria for the demand: the positive (additive) externalities in a market context may give raise to a family of non-monotonic demand curves generalizing thus the classical ones. 

Beyond this first main result, we exhibit the generic properties of the boundaries limiting the regions 
where the system presents different types of equilibria (unique or multiple).
We call these properties {\em generic} since we show that they depend {\em only} on qualitative features of the 
IWP distribution: modality (number of maxima), smoothness (continuity and derivability properties) 
and type of support (compact or infinite).
The main results are summarized as {\em phase diagrams} in the space of the relevant parameters of the model, namely (i) the social influence strength and (ii) the difference between the population average of the IWP and the posted price, both parameters conveniently normalized by the standard deviation of the IWP distribution (which measures the importance of the heterogeneity).

\subsection{Organization of the paper}
\label{sec:org}
The paper is organized as follows. In Section \ref{sec:model} we present the model: we first (section \ref{sec:customer}) specify the agents (customers) model, then in \ref{sec:basic_equations}
we introduce a normalized form of the basic equations
which is convenient for analyzing the demand, and in \ref{sec:hints} 
we show on two simple extreme cases what to expect from these equations.
In \ref{sec:hypotheses} we detail the families of probability distributions of the IWP covered by this paper.

In Section \ref{sec:demand} we analyze the aggregate demand (its collective behavior)
for a generic smooth pdf. In \ref{sec:directinverse} we introduce and study the
direct and inverse demand functions; the demand phase diagram is derived in Section \ref{sec:demand_msr}: in \ref{sec:boundary}
we obtain the domain of multiple solutions which allows to plot the 
phase diagram (section \ref{sec:demand_phasediag}). We analyze in details the vicinity of
the bifurcation point (section \ref{sec:demand_B}) and the question of Pareto optimality (section \ref{sec:Pareto}). A summary of the demand properties is given in \ref{sec:demand-summary}.

We leave to the Appendix \ref{app:other_distr} the analysis of other kinds of pdfs, where
we study the demand phase diagrams of 
IWP distributions with compact support (section \ref{sec:compactsupport}) and with
fat tails (section \ref{sec:fat}). The demand phase diagram for a pdf with an arbitrary number of maxima is studied in \ref{sec:app_multimodal} ---in details for a smooth multimodal pdf in Section 
\ref{sec:app_multimodal_smooth}, and on a simple example of singular bimodal distribution in
section \ref{sec:app_2diracs}---. 

Finally we summarize the main results and give several perspectives for further research in Section \ref{sec:conclusion}.

\section{Model of discrete choices with heterogeneous agents and positive externalities}
\label{sec:model}

\subsection{Agents model}
\label{sec:customer}
We consider a population of $N$ agents ($i=1,2, \dots, N$). Each individual $i$ has to make
a binary choice. 
Depending on the context, this binary decision may represent the fact of buying or not a good, 
adopting or not a given standard, 
adopting or not some social behavior such as joining a 
riot \cite{Granovetter78}, or a journal club \cite{Schelling73}, \cite{Schelling}, etc. 
Formally each agent $i$ must choose a strategy $\omega_i$ in the strategic set 
$\Omega = \{0,1\}$ ($\omega_i=1$ denotes to buy/adopt/join, $\omega_i=0$ otherwise)\footnote{Some authors use the notation $s_i=1$ and $s_i=-1$; both encodings are equivalent: 
it suffices to replace $\omega_i = (s_i+1)/2$ in our model and identify the coefficients 
of corresponding expressions.}. Hereafter, without loss of generality,
we will refer to the simplest market situation where the agents are customers who must choose  
whether to buy or not a single good at a price $P$. Our main concern is with the agents' behaviors,
and $P$ is considered as an exogenous parameter ---e.g. it is
posted by a monopolist selling the good---. Non-market models like 
those recently considered by, i.e., Glaeser et al. \cite{GlaeserScheinkman} are obtained by setting $P=0$ 
or by considering $P$ as an exogenous social cost, common to all the agents.
We are interested in the collective outcome of the agents decisions.  
In a following paper \cite{mono} focusing on the market context we will 
analyze the consequences of the customers collective behavior
on a monopolist's program for fixing the optimal price. 

The population is heterogeneous. Each individual $i$ has an {\em idiosyncratic} preference 
or {\em willingness to pay/adopt} (hereafter IWP) $H_i$, meaning that in the absence of social influences, an agent $i$ adopts the state $\omega_i=1$ if $H_i$ is larger than the price $P$. 
Following Mc-Fadden~\cite{McFadden74} and Manski \cite{Manski77}, we work within the framework of
{\em Random Utility Models} (RUM):
we assume that the $H_i$ are time independent random variables independently and identically distributed (i.i.d.) 
in the population. Denoting by $H$ the mean and by $\sigma$ the variance of the IWP distribution,
hereafter we assume that the random variable $(H_i - H)/\sigma$ is distributed according to:
\begin{equation}
{\mathcal P}( x < \frac{H_i - H}{\sigma} < x + dx ) 
= f(x) dx,
\label{eq:pdf}
\end{equation}
so that $f$ is a pdf with zero mean and unitary variance.
In the RUM view point, the agents have utilities $H_i^0$ and $H_i^1$ for not adopting and adopting
respectively, with $H^0$ and $H^1$ the corresponding population averages.
For $k=0,1$, one writes $H_i^k= H^k + \epsilon_i^k$, where
$\epsilon_i^k$ are independent random variables with zero mean and standard deviation $\sigma^k$,
with not necessarily  identical pdfs for $k=0$ and $k=1$. Then in our model
we have $H_i=H_i^1-H_i^0$, $H=H^1-H^0$ and $f$ is the pdf of the normalized difference
$x_i= (\epsilon_i^1 - \epsilon_i^0)/\sigma$,
with $\sigma^2=(\sigma^0)^2+(\sigma^1)^2$ because of the additivity of the variances of
independent variables. As particular examples, if $\epsilon_i^k$ are Gaussian variables,
then $x_i$ is also Gaussian; if  $\epsilon_i^k$ are uniformly distributed on, say
the intervals $[-a^k, a^k]$, then $x_i$ has a trapezoidal pdf, which becomes a triangular distribution
if e.g. $a^0=0$ (no uncertainty in the utility of not adopting). In the following we do not
assume any specific form of the pdf of $x$. 
In Section \ref{sec:hypotheses} below, 
we present in details the class of pdfs considered in this paper. 

If all the individuals had the same IWP, the outcome in absence of social interactions would be very simple: either the price is below this common value, and everybody buys, or it is above it and nobody buys. All the individuals would behave in the same way, obtaining the same payoffs, and in the market aggregate analysis, they may be replaced by a fictitious {\it representative agent} \cite{Kirman92}. In the case of a 
heterogeneous population considered here, only the agents with $H_i \geq P$ would buy at price $P$, but getting different payoffs. 

The situation is more complex when the decision of each agent depends {\it also} on the decisions of others (\cite{Manski77} and references therein). We assume that each agent is the more willing to pay the larger the number of buyers in the population. We consider a linear separable surplus, that is {\it if} agent $i$ buys at the posted price $P$, his surplus is
\begin{equation}
S_i =  H_i + J \eta - P,
\label{eq:u_i}
\end{equation}
where $\eta$ is the fraction of buyers in the population. Taking into account the definition of $\omega_i$: 
\begin{equation}
\eta \equiv \frac{1}{N} \sum_{i=1}^N \omega_i.
\label{eq:eta}
\end{equation}
We assume also that the externality $J \eta$ corresponds to strategic complementarities, i.e. that the strength of the social influence is positive: $J > 0$\hspace{1pt}
\footnote{More generally, the social term may be proportional to the fraction of buyers in an individual-depending subset of the population, called ``neighbors'' of agent $i$. In this paper, we consider a global neighborhood, where every agent has social connections with every other agent, mainly because this case can be studied analytically.}.

The actual surplus of agent $i$ is: 
\begin{equation}
W_i = S_i \; \omega_i.
\label{eq:payoff}
\end{equation} 
In order to maximize his surplus, agent $i$ should buy/adopt ($\omega_i=1$) if $S_i > 0$, but not ($\omega_i=0$) when $S_i<0$. Since the IWP are i.i.d., when $N$ is very large (more precisely, in the limit $N \rightarrow \infty$), by the law of large numbers, the fraction of buyers (\ref{eq:eta}) ---which is the average of $\omega_i$--- converges to the expected value of $\omega_i$ over the IWP distribution. Thus, $\eta$ is given by the fixed point equation:
\begin{equation}
\label{eq:fp1}
\eta = {\mathcal P}(H_i - P + J \eta > 0).
\end{equation}
The marginal customer $m$, indifferent between adopting or not, is defined by the condition of zero surplus, $S_m=0$:
\begin{equation}
H_m - P + J \eta = 0
\label{eq:Sm}
\end{equation}
so that (\ref{eq:fp1}) may be written as
\begin{equation}
\label{eq:fpm}
\eta = {\mathcal P}(H_i > H_m).
\end{equation}
For what follows it is more useful to write (\ref{eq:fp1}) as
\begin{equation}
\label{eq:fp2}
\eta = {\mathcal P}(H_i - H > -S)
\end{equation}
where
\begin{equation}
S = S(J,H,P;\; \eta) \equiv H - P + J \eta
\label{eq:mean_s}
\end{equation}
is the population average of the ({\em ex ante}) surplus $S_i$. 
It depends on the parameters $J$ and $H$, which are properties of the customers population, and $P$, the exogenous price. 

\paragraph{Notation.} Generally, 
upon manipulating functions, we put in parenthesis the parameters in front, separated with a semicolon ($;$) from what we consider the variable. Our notation $S(J,H,P;\eta)$ in (\ref{eq:mean_s}) indicates 
that $J$, $H$ and $P$ are considered as parameters, whereas $\eta$ is the variable. Sometimes, when the context is clear, we drop down the parameters and keep only the variable, writing thus $S(\eta)$. Whenever we consider functions of two variables, they are separated by a simple colon, like in equations (\ref{eq:pk}) and (\ref{eq:slope}), and in Appendix \ref{sec:app_multimodal}.

\subsection{Aggregate behavior: normalized equations}
\label{sec:basic_equations}
Clearly, the fraction of buyers $\eta$ depends on the strength of the social 
influence $J$, the price $P$ and the average willingness to pay in the population $H$, 
and on the distribution of the deviation of the IWP $H_i$ from its population average $H$. 
The agents choices depend only on the surplus sign, and they are invariant under changes of 
the surplus scale. Since the surplus is linear, we can formally multiply every term of the surplus by a same 
strictly positive number without changing the agents' choices. An adequate scale is given by the typical scale 
of the IWP distribution: it is convenient to measure each quantity ($J, H, P$) in units of the width $\sigma$ 
of the IWP pdf. Hence instead of four parameters, we are left with three independent parameters.

Hereafter we will thus work with the following normalized variables
\begin{equation}
j \equiv \frac{J}{\sigma}, \;\;\;\;
h \equiv \frac{H}{\sigma}, \;\;\;\;
p \equiv \frac{P}{\sigma}
\label{eq:reduced_var}
\end{equation}

In addition, as it is obvious from equations (\ref{eq:fp2}) and (\ref{eq:mean_s}), $\eta$ depends on 
the price $P$ and the average willingness to pay $H$ only through their difference $H-P$. 
We introduce the normalized difference:
\begin{equation}
\delta \equiv \frac{H - P}{\sigma} \; = h - p,  
\label{eq:delta}
\end{equation}
which is the average ex-ante surplus in the absence of externality. For short hereafter we call $\delta$ 
the {\em bare surplus}. 
In non-market models ($p=0$) it is the  
average willingness to adopt.

\paragraph{Remark.} In (almost) all the following we will work with the above reduced variables (\ref{eq:reduced_var}), (\ref{eq:delta}),
referring to them as the (normalized) strength of social influence, average willingness to pay, price,
and bare surplus. However one should keep in mind, especially when interpreting the results, 
that they represent the {\em ratios} of the non normalized parameters to the width of the IWP distribution. Clearly other normalizations are
possible. An alternative of particular interest is the normalization obtained by
measuring every quantity in units of the social strength $J$: the relevant parameters are then
\begin{equation}
\tilde \sigma \equiv \frac{\sigma}{J}, \;\;\;\;
\tilde h \equiv \frac{H}{J}, \;\;\;\;
\tilde p \equiv \frac{P}{J}, \;\;\;\;
\tilde \delta \equiv \frac{H - P}{J} \; = \tilde h - \tilde p
\label{eq:reduced_var_bis}
\end{equation}
(equivalently one can do as if $J=1$). 
Note that this choice of 
normalization is no more than an equivalent representation of the
parameters space; indeed one has $ \tilde \sigma =1/j,\;\;\;\tilde h = h/j,\;\;\; \tilde p = p/j$. 
It is also interesting to analyze the results in term of the
set of parameters (\ref{eq:reduced_var_bis}), which allow to stress the model
properties as a function of the degree of heterogeneity (relative to the strength of
the social influence) ---a homogeneous population
corresponding to the limiting case $\tilde \sigma = 0$, a highly heterogeneous one to a large 
$\tilde \sigma$---. 
 
\vspace{1cm}
 
With the normalized variables (\ref{eq:reduced_var}), (\ref{eq:delta}), equation (\ref{eq:fp2}) becomes
\begin{equation}
\eta =\int_{-s}^\infty f(x) dx = 1 - F(- s),
\label{eq:fp}
\end{equation}
where $F$ is the cumulative probability distribution and $s =  S/\sigma$, with $S$ defined by (\ref{eq:mean_s}), depends on $h$ and $p$ through the bare surplus $\delta$, that is 
\begin{equation}
s = s(j,\delta;\eta) \equiv \delta + j \eta.
\label{eq:s} 
\end{equation}

If the pdf has infinite support,  
\begin{equation}
F(-s)\equiv {\mathcal P}(x \leq -s) = \int_{-\infty}^{-s} f(x) dx.
\label{eq:F}
\end{equation}
In the case of a compact support $[x_m,\; x_M]$, one can write:
\begin{equation}
\eta=1-F(-s) = \int_{\max\{x_m, -s\}}^{\max\{x_M, -s\}} f(x) dx.
\label{eq:fpb}
\end{equation}
Obviously, when $-s<x_m$, we have $\eta=1$, and when $-s>x_M$ we have $\eta=0$.

\subsection{Hints from two extreme cases}
\label{sec:hints}
In the absence of social influence the problem is simple because after introduction 
of $j=0$ in (\ref{eq:s}) we obtain $s=\delta$ which does not depend on $\eta$. 
Then, due to the monotonicity of cumulative distributions, the fraction of buyers 
(\ref{eq:fp}) is a monotonically increasing function of $\delta$ 
(equivalently, at fixed $h$, a decreasing function of the price $p$):
\begin{equation}
\eta=1-F(- \delta).
\label{eq:eta_j0}
\end{equation}

Another extreme case is that of a homogeneous population: $H_i=H$ for every $i$
---a situation obtained in the singular limit $\tilde \sigma = 1/j \rightarrow 0$ (the
IWP distribution becoming a Dirac distribution)---. In that case every agent is faced to exactly the same
decision problem, so that at equilibrium either $\eta=0$ or $\eta=1$.
For each agent the surplus in case of adoption would be $H -P$ if no other agent adopt ($\eta=0$), whereas if 
$\eta=1$ the surplus is $H-P + J$. In fact, $\eta=0$ is a solution for $H < P$, while $\eta=1$ is a solution 
for $H > P - J$. Hence there is a domain, $P-J < H < P$, where the two solutions coexist.
The whole population behaves as a single
agent who either does not adopt, $\eta=0$, or adopts, $\eta=1$, with a different surplus depending on whether he
is ``in" or ``out of" the market: this is analogous
to the problem of multi-equilibria with hysteresis in trade analyzed by Baldwin and Krugman \cite{BaldwinKrugman},
except that here the problem arises only at the {\em collective level}.

We have thus on one side, for $J=0$ and $\sigma$ finite, a unique well behaved equilibrium, and for
$J > 0$ but $ \sigma=0$ a situation of multiequilibria. The question addressed in the following aims at understanding
what happens ``in between".
We will show that when the social interaction strength is large enough compared to the
heterogeneity width, the demand faces a complex problem. More precisely, there is a critical 
value $j_B$ of $j=J/\sigma$. Below it, the fraction of buyers (equation (\ref{eq:fp})) follows monotonically the price variations. Beyond $j_B$, equation (\ref{eq:fp}) presents multiple solutions. 
Among them, the (possibly multiple) Nash equilibria are those solutions that have an economic meaning, i.e. 
for which the demand decreases when prices increase.

The Section \ref{sec:demand} of the paper is devoted to a detailed study of the nature of the solutions of 
equations (\ref{eq:fp}) and (\ref{eq:s}) with $j \geq 0$ for distributions satisfying very general smoothness 
hypotheses, detailed in the next section.

\subsection{The idiosyncratic willingness-to-pay distribution}
\label{sec:hypotheses}
Since we are interested in the generic properties of the model, we explicit the general characteristics of the idiosyncratic willingness-to-pay (IWP) distributions covered by our analysis.

Since a pdf must be integrable, $f(x)$ (equation (\ref{eq:pdf}) ) must vanish in the limits $x \rightarrow \pm \infty$. For sufficiently regular pdfs, this can happen in two different ways: either the pdf decreases continuously to $0$ as $x \rightarrow \pm \infty$, or it is strictly zero outside some compact support $[x_m, x_M]$. Most of the analysis in this paper is restricted to the class of pdfs obeying to the following hypotheses:

\begin{enumerate}
\item[H1.] {\em Modality}: $f$ is {\em unimodal}, that is it has a {\em unique} maximum 
\begin{equation}
f_B \equiv \sup_{x} f(x).
\label{eq:fB}
\end{equation} 

\item[H2.] {\em Smoothness}: $f$ is non zero, continuous, and at least piecewise twice continuously differentiable inside 
its support, $]x_m, x_M[$ , where $x_m$ and $x_M$  may  be finite or equal to $\pm \infty$. 
In the latter case $f$ is stricly monotonically decreasing towards zero 
as $x \rightarrow \pm \infty$. 

\item[H3.] {\em Boundedness}: the maximum of $f$, $f_B$ (that may be reached at $x_m$ or $x_M$ if these numbers are finite), is {\em finite}:
\begin{equation}
f_B < \infty
\label{eq:fB_finite}
\end{equation} 
\end{enumerate}

Within the class of pdfs satisfying H1, H2 and H3, we will consider more specifically the important following prototypical cases:
\begin{enumerate}
\item [1.] {\em Unbounded supports}: The support of the distribution is the real axis; 
the pdf is continuous and twice continuously derivable on $]-\infty, \infty[$, with a unique maximum. A typical example, relevant to economics (see e.g. \cite{AnDeTh}), is given by the logistic distribution, although more generally we do not assume that the pdf is symmetric. 
We make the following supplementary hypothesis, that amount to impose that the pdf decreases fast enough for $x \rightarrow \pm \infty$:
\begin{enumerate}
\item[H4.] {\em Mean value}: the pdf has a finite mean value. Then, the smoothness condition H2 imposes that $f$ decreases when $x \rightarrow \pm \infty$ faster than $|x|^{-1}$. 
\item[H5.] {\em Variance}: the pdf has a finite variance.  
Then, the smoothness condition H2 imposes that $f$ decreases when $x \rightarrow \pm \infty$ 
faster than $|x|^{-2}$. 
\end{enumerate}
\item [2.] {\em Compact supports}: the support of the distribution is some interval $[x_m, x_M]$, with $x_m$ and $x_M$ finite; the pdf is continuous on $[x_m, x_M]$ and continuously derivable on $]x_m, x_M[$, with a unique maximum on $[x_m, x_M]$. Note that, since $f$ has zero mean, $x_m < 0 < x_M$. 
\end{enumerate}

Hypothesis H2 and H3 exclude cases where the pdf is not a function but a distribution --- containing, e.g., a Dirac delta ---. Clearly, if the pdf's support is the real line, $]-\infty, +\infty[$, the boundedness hypothesis H3 is a consequence of 
the smoothness hypothesis H2. In the case of compact supports, H3 excludes pdfs diverging at a boundary of the support. Although hypothesis H3 is actually true under H2 if $f$ is continuous on the closed interval $[x_m, x_M]$, we explicit it because some of our results are valid under H3 even for pdfs less regular than those satisfying H2. 

Although hypothesis H5 is not necessary for the study of the aggregate demand, it corresponds to a wide family of realistic distributions for which one can conveniently use the standard deviation as the unit for measuring the relevant parameters (i.e., using normalization (\ref{eq:reduced_var})).

Generic results for unbounded support pdfs satisfying H1 to H5 are presented in the main body of the paper. 
They extend previous results obtained for a logistic distribution \cite{NaPhGoVa05}. The analysis of other types of pdfs
 is left to Appendix \ref{app:other_distr}:
\begin{itemize}

\item[-] In Appendix \ref{sec:compactsupport} we present general results for bounded support pdfs. The case of a uniform distribution on a 
finite interval $[x_m, x_M]$, which corresponds to an interesting degenerate case ($f$ is maximal at every point within the 
interval), has been presented elsewhere \cite{GoNaPhVa05}. The particular case of general a triangular pdf is explicitly worked out for illustration.

\item[-]In Appendix \ref{sec:fat} we extend the analysis to {\em fat-tail} distributions, which correspond to an important limiting case 
of pdfs with infinite variance (for such distributions, the normalization constant $\sigma$ in (\ref{eq:pdf}) 
and in equations (\ref{eq:reduced_var}) 
and (\ref{eq:delta}) is no longer the standard deviation, but an arbitrary positive constant setting the units of $H$, $J$, $P$ and $C$). 

\item[-] Finally, in Appendix \ref{sec:app_multimodal} we extend the discussion to the case of multimodal pdfs 
(distributions with an arbitrary number of maxima): we derive the demand phase diagram for a generic smooth multimodal pdf 
and we discuss  
the case of a singular pdf using as an illustrative example a pdf with two Dirac peaks.
\end{itemize}

\section{Aggregate choices and coordination dilemma}
\label{sec:demand}

In this section we discuss the demand function, that is the relationship
between price $p$ and fraction of buyers (or adopters, in non market contexts) $\eta$, expressed by equation (\ref{eq:fp}). As we have already seen, this means studying the relationship between $\eta$ and the bare surplus $\delta = h-p$, and how it depends on the externality parameter $j$. 
We show that, for a large range of values of the parameters $j$ and $\delta$, the demand presents two 
equilibria which can be qualified as Nash equilibria from a game-theoretic point of view. 
This result is valid for any pdf satisfying the general hypothesis described in the preceding section.

\subsection{The direct and inverse demand functions}
\label{sec:directinverse}

The expected demand $\eta^d$ at a given value of $\delta$  
is obtained as the implicit solution of (\ref{eq:fp}) and (\ref{eq:s}). 
As we will see, the application $\eta \rightarrow 1 - F(-s(\eta))$ may be a multiply valued function of $\eta$; 
it is thus preferable to express $\delta$, or $p=h-\delta$, as a function of $\eta$, and determine the 
{\em inverse demand function} $p^d(\eta)$, that is, the price at which exactly $N \eta$ units of the good 
would be bought\footnote{In non market models, where generally $p=0$, results in this section give the aggregate choice
$N \eta^d(j,h)$ as a function of $j$ and $h$. This is the relationship between the fraction of adopters, the average willingness to adopt of the population 
and the strength of the social interactions.}. 

\begin{figure}
\centering
\includegraphics[width=0.7\textwidth]{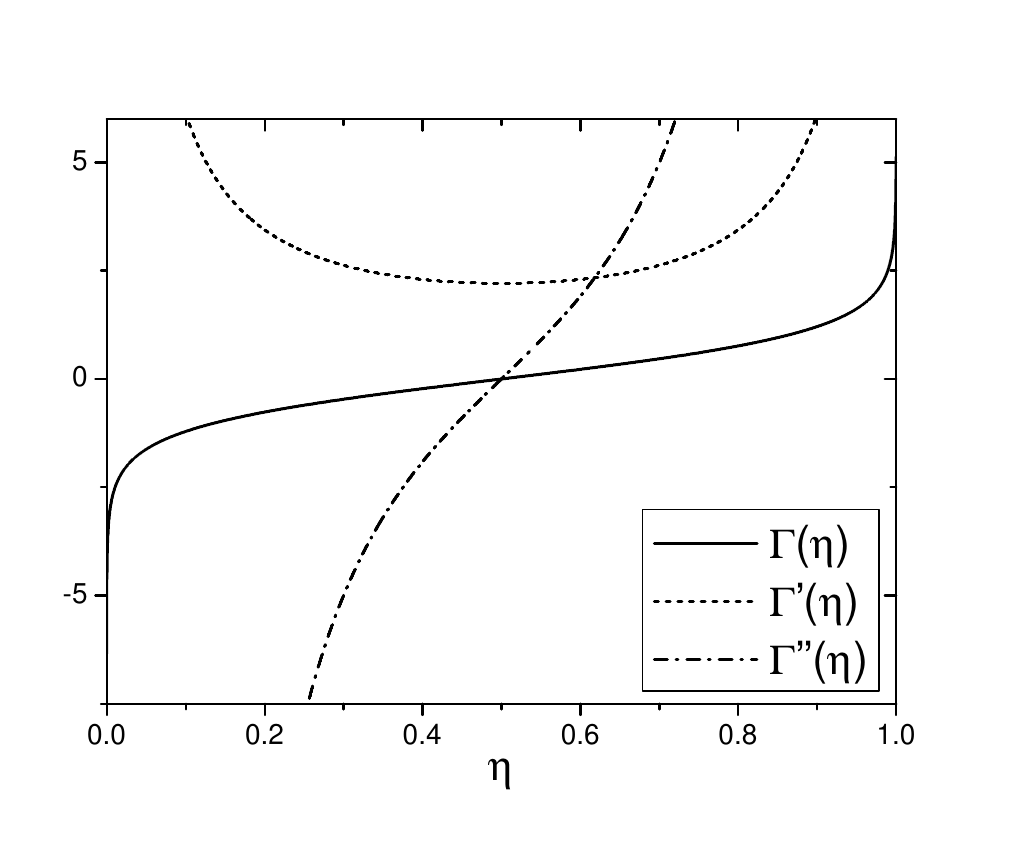}
\caption[]{\em $\Gamma(\eta)$ and derivatives as a function of $\eta$ for the logistic pdf of unitary variance. Remark: all these functions diverge at $\eta=0$ and at $\eta=1$.}
\label{logistic.fig}
\end{figure}

Under the hypothesis H2 the cumulative distribution $F$ is a continuous and {\em strictly} 
monotonic function on $]x_m, x_M[$ and has a unique inflexion point. Hence it is invertible. 
Denoting $\Gamma$ the inverse of $1-F(-s)$, we have the following equivalence:
\begin{equation}
\eta = 1-F(-s)  \; \Longleftrightarrow \; s = \Gamma(\eta),
\label{eq:Gamma} 
\end{equation}
with $s$ defined by (\ref{eq:s}). For unbounded supports, $\Gamma(\eta)$ increases monotonically 
from $- \infty$ to $+\infty$ when $\eta$ goes from $0$ to $1$ (see figure \ref{logistic.fig} for an example, and Appendix \ref{app:other_distr} for other cases). In the case of a compact support $[x_m,x_M]$, $\Gamma(\eta)$ takes the finite values, $\Gamma(0) = -x_M$, and $\Gamma(1) = -x_m$ for $\eta$ respectively 
$0$ and $1$. Note that neither we assume $f$ to be symmetric nor to have its maximum at $x=0$.

Replacing $s$ in the r.h.s. of (\ref{eq:Gamma}) by its expression (\ref{eq:s}) yields
\begin{equation}
\delta = {\cal D}(j;\eta), 
\label{eq:h-p} 
\end{equation}
with 
\begin{equation}
{\cal D}(j;\eta) \equiv \Gamma(\eta)-j \eta. 
\label{eq:D} 
\end{equation}
Interestingly, ${\cal D}(j;\eta)$ depends on the parameter $j$ but not on $h$. 
Actually, in Section \ref{sec:demand_msr} and in Appendix \ref{sec:app_multimodal}, 
we will have to consider ${\cal D}(j;\eta)$ 
as a function of the two variables, $j$ and $\eta$. 
In the present subsection however, we consider ${\cal D}(j;\eta)$ as a function of the single variable $\eta$, 
with $j$ as a (fixed) parameter - hence according to our convention on notations introduced in Section \ref{sec:customer},
we keep in mind the dependency of ${\cal D}$ on $j$ by writing ${\cal D}(j;\eta)$, and 
derivatives of ${\cal D}(j;\eta)$ with respect to $\eta$ are denoted ${\cal D}'(j;\eta)$. 

\begin{figure}
\centering
\includegraphics[width=0.7\textwidth]{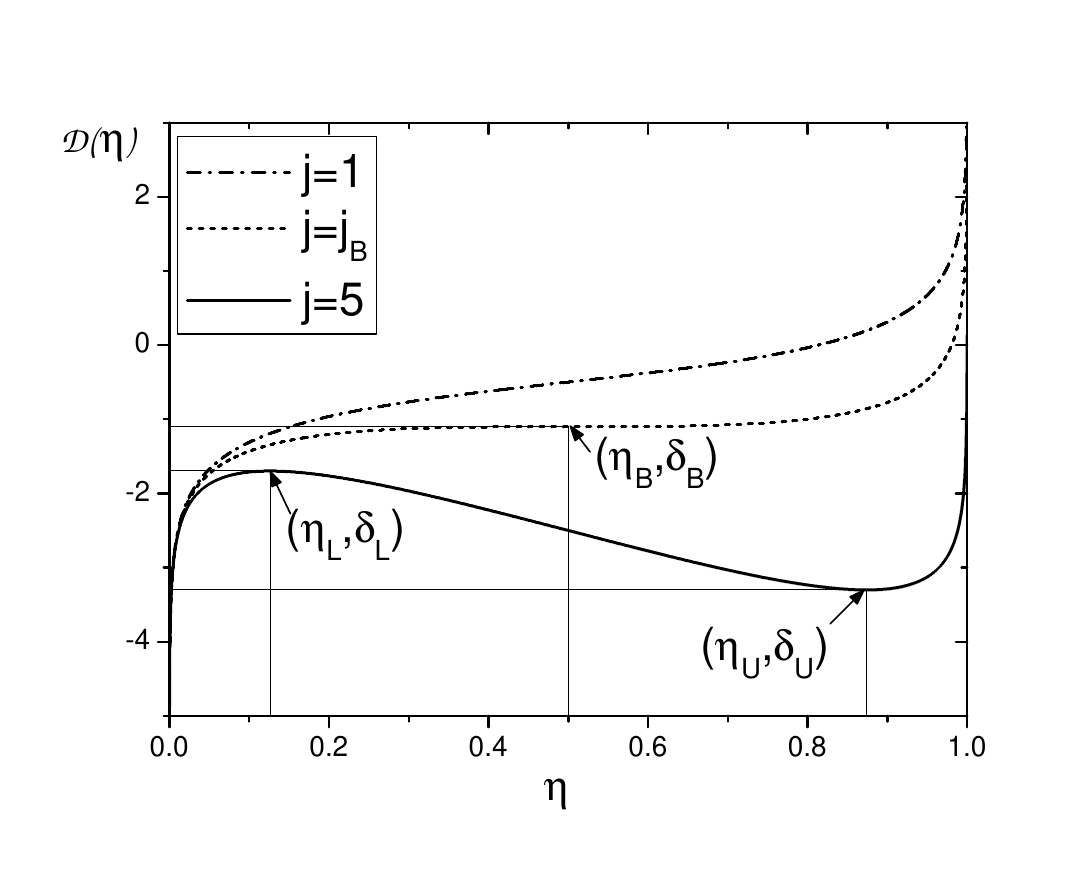}
\caption[]{\em {${\cal D}(j; \eta)$ as a function of $\eta$ in the case of a logistic distribution of the IWP, for three values of $j$: $j=1<j_B$, $j=j_B=2.20532$ and $j=5 > j_B$.}}
\label{fig:calD}
\end{figure}

Plots of ${\cal D}(j;\eta)$ against $\eta$ for different values of $j$ are presented on figure \ref{fig:calD} 
for the logistic distribution. Solutions to (\ref{eq:h-p}) correspond to the intersections 
of these functions with horizontal lines at $y=\delta$.

The stable equilibrium values of the demand satisfy
\begin{equation}
{\cal D}'(j; \eta) \ge 0.
\label{eq:stability_cond}
\end{equation} 
Thus, the intersections of $y=\delta$ with $\cal D$ when ${\cal D}'<0$ correspond to unstable equilibria and will be ignored, as explained below.

Given an externality parameter $j$ and a given value of $\delta$, the solution $\eta=\eta^d(j;\delta)$ of equation (\ref{eq:h-p}) satisfying (\ref{eq:stability_cond}) gives the expected demand $N\eta^d(j;\delta)$ at a price $p = h-\delta$. 

From the definition of $\delta$ and (\ref{eq:D}), the inverse demand function is thus 
\begin{equation}
p^d(\eta) = h - {\cal D}(j; \eta).
\label{eq:p^d}
\end{equation}
This function depends on both parameters $h$ and $j$, and when necessary we will write $p^d(\eta) =p^d(h,j; \eta)$. 

As we will see, the demand $\eta^d(j;\delta)$ can be a multivalued function of $\delta$ for some range of 
parameters. On the contrary, since the function $\Gamma(\eta)$ is a uniquely defined function of 
$\eta$, so is ${\cal D}(j; \eta)$. This is the reason why, instead of considering (\ref{eq:fp}), we prefer to obtain the properties of the demand $\eta^d(j;\delta)$ from the analysis of equations (\ref{eq:h-p}) and (\ref{eq:D}). 

Under assumption H2, $\Gamma(\eta)$ is at least piecewise three times continuously derivable on $]0, 1[$; 
its derivative $\Gamma'$ is continuous and strictly positive. In particular, at any point $\eta$ in $]0,1[$, we have:
\begin{equation}
{\cal D}'(j; \eta) = \Gamma'(\eta) - j.
\label{eq:dD}
\end{equation}
In the case of a compact support, the above equation also holds for the right and left derivatives at, 
respectively, $\eta=0$ and $\eta=1$. 

In terms of the pdf $f$,
\begin{equation}
\Gamma'(\eta) = \frac{1}{f(-s)} \;\mbox{, with} \;\; s=\Gamma(\eta).
\label{eq:dG/dn}
\end{equation}
Under H1, $\Gamma'$ has a unique absolute minimum (qualitatively there is a unique point where
the curvature of $\Gamma$ changes from convex to concave; if $\Gamma$ is smooth, it has a unique 
inflexion point). Thus
\begin{equation}
\min_{\eta} \Gamma'(\eta) = \frac{1}{f_B} \; > 0.
\label{eq:minG'}
\end{equation}
This minimum is reached at some value $\eta=\eta_B$: 
\begin{equation}
\eta_B \equiv \arg \min_{\eta} \Gamma'(\eta).
\label{eq:etaB}
\end{equation}
If $f$ is smooth enough at its maximum, then $\Gamma''(\eta_B)=0$: $\eta_B$ is the inflexion 
point of $\Gamma$. For symmetric pdfs, $\eta_B=1/2$, but we do not restrict to this case.

As a consequence of the properties of $\Gamma(\eta)$, we see from equation (\ref{eq:dD}) that ${\cal D}'(j; \eta)$ is strictly positive for $j < j_B$, with
\begin{equation}
j_B \equiv \Gamma'(\eta_B) = \frac{1}{f_B}.
\label{eq:jB}
\end{equation}

The value $j_B$ separates two regions where the model presents qualitatively different behaviors. When $j<j_B$, 
the function ${\cal D}(j; \eta)$ is strictly increasing from $- \infty$ to $+\infty$ as $\eta$ goes from $0$ to $1$. 
As a result it is invertible: for any $\delta$ in $]- \infty, +\infty[$, equation (\ref{eq:h-p}) 
has a unique solution $\eta^d(\delta)$. 

If $j > j_B$, there is a range of values of $\delta$ for which (\ref{eq:h-p}) has several solutions. 

The existence of {\em multiple solutions} in the demand is thus a {\em generic property of discrete choice models with 
heterogeneous agents and social interactions} (externalities). This is true whatever the number of maxima of $f$, as shown in Section \ref{sec:app_multimodal} of Appendix \ref{app:other_distr}. 
Actually, the domain where there is a unique solution, that is $0 \leq j \leq j_B=1/f_B$, is very narrow if $f_B$ is large: 
a leptokurtic distribution 
will have in general a narrower domain of unicity of the demand than a platykurtic distribution of same variance.

\begin{figure}
\centering
\includegraphics[width=0.7\textwidth]{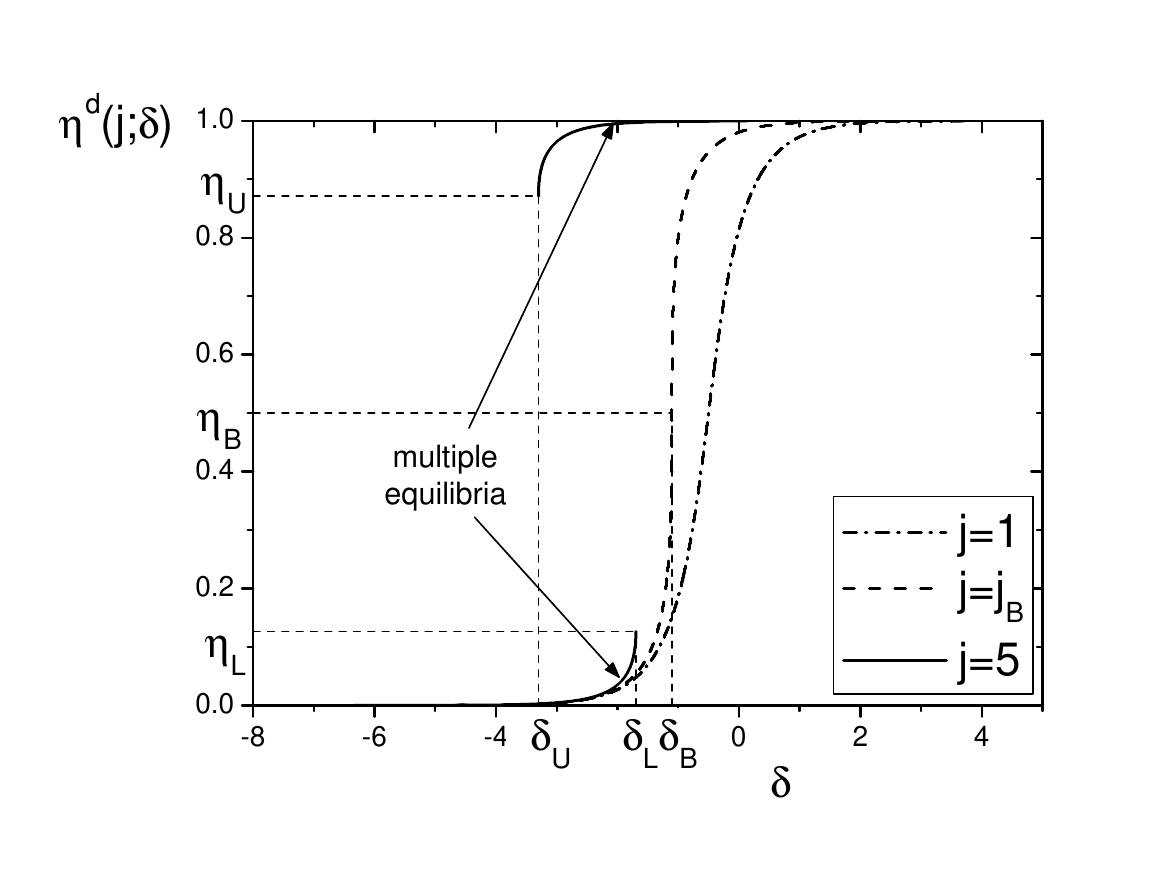}
\caption[]{\em {Demand $\eta^d(j;\delta)$ as function of $\delta \equiv h-p$ in the case of a logistic distribution of the IWP, for three values of $j$: $j=1<j_B$, $j=j_B=2.20532$ and $j=5> j_B$. Notice that the origin of the horizontal axis ($\delta=0$) corresponds to $h=p$. Remark that prices {\em increase} from {\em right to left}. Unstable solutions: the negative slope curve joining $(\delta_L(j),\eta_L(j))$ to $(\delta_U(j),\eta_U(j))$ for $j=5$ is not shown.}}
\label{fig:EtaOfDelta}
\end{figure}

In our case of unimodal pdfs, equation (\ref{eq:h-p}) may have three solutions for
$j > j_B $ (see figure \ref{fig:calD}). The intermediate solution, laying on a branch with ${\cal D}'(j; \eta) < 0$ 
---where $\eta$ increases as $\delta=h-p$ decreases--- is sometimes called a {\em critical mass} 
point in the literature \cite{Schelling73}: it corresponds to a demand that 
would increase for increasing prices. Hence, in a {\em tatonnement} dynamics, this corresponds 
to an unstable solution separating the basins of attraction of the two stable equilibria. 
The marginal case, $j=j_B$, is a bifurcation point (hence the subscript $B$) where multiple 
solutions to (\ref{eq:h-p}) appear on increasing $j$. The stable equilibria of the demand that satisfy (\ref{eq:stability_cond}) are represented against $\delta$ on figure \ref{fig:EtaOfDelta}.

\subsection{Demand phase diagram}
\label{sec:demand_msr}

\subsubsection{The demand multiple-solution region}
\label{sec:boundary}
Let us consider more in details the behavior of the application $\delta \rightarrow \eta^d(\delta)$ in the case of a smooth unimodal pdf on $]-\infty, +\infty[$. Considerations specific to compact support pdfs are left to Section \ref{sec:compactsupport} of Appendix \ref{app:other_distr}.

The functions $\Gamma(\eta)$ corresponding to pdfs satisfying H1 to H5 are at 
least three times continuously derivable on $]0,1[$, and diverge towards $-\infty$ and $+\infty$ as $\eta$ goes to $0$ and $1$ respectively. We have already seen that for $j < j_B$ there is a unique solution, and $\eta^d$ goes from $0$ to $1$ as $\delta=h-p$ goes smoothly from $-\infty$ to $+\infty$. 

For $j> j_B$, (\ref{eq:h-p}) has 3 solutions whenever $\delta_U < \delta < \delta_L$ (see figure \ref{fig:calD}), where $\delta_L$ and $\delta_U$ are the values of $\delta$ that satisfy the equality (marginal stability condition) in equation (\ref{eq:stability_cond}). That is, the boundaries of the region with multiple solutions are the values for which ${\cal D}(j; \eta)$ has a horizontal slope (see figure \ref{fig:calD}):
\begin{equation}
{\cal D}'(j; \eta)= 0
\label{eq:jdmargin=0}
\end{equation} 
which is equivalent to 
\begin{equation}
\frac{ dp^d(\eta) }{ d\eta }=0.
\label{eq:dpd=0}
\end{equation}
Considering the definition (\ref{eq:D}) of ${\cal D}$, this means that on these boundaries 
${\cal D} (j,\eta(j))$, {\em as a function of $j$}, is the Legendre transform of $\Gamma(\eta)$.  
Under our hypothesis H1, $\Gamma'$ has a unique minimum, and necessarily tends towards $+\infty$ as $\eta$ 
goes to either $0$ or $1$; $\Gamma$ is strictly convex on $]\eta_B, 1[$, and strictly concave on $]0, \eta_B[$, hence the Legendre transform is well defined and unique on each one of these intervals:
equation (\ref{eq:jdmargin=0}) for $j> j_B$ has indeed two solutions $\eta_L(j)$ and $\eta_U(j)$, given by
\begin{equation}
j = \Gamma'(\eta_\Lambda) , \;\; \Lambda=U,L.
\label{eq:jdmargin}
\end{equation}
with
\begin{equation}
\eta_L(j) < \eta_B < \eta_U(j).
\label{eq:eta12}
\end{equation}

From the knowledge of $\eta_U(j)$ and $\eta_L(j)$, using (\ref{eq:h-p}) one gets the marginal stability curves $\delta_U(j)$ and $\delta_L(j)$, that is, the extreme values 
of $\delta$ bounding the region where multiple solutions exist: 
\begin{equation}
\delta_\Lambda(j) =  {\cal D}(j ; \eta_\Lambda(j)) = \Gamma(\eta_\Lambda(j)) - j \eta_\Lambda(j) , \;\; \Lambda=U,L.
\label{eq:delta1delta2}
\end{equation}

As already stated, for $\delta_U(j) < \delta < \delta_L(j)$, equation (\ref{eq:h-p}) has three solutions. The curve $\eta^d(j;\delta)$ has two stable branches (see figure \ref{fig:EtaOfDelta}): an upper one $\eta_U^d(j;\delta)$ with $\eta_U^d(j;\delta) > \eta_U(j) > \eta_B$, and a lower one $\eta_L^d(j;\delta)$ with $\eta_L^d(j;\delta) < \eta_L(j) < \eta_B$; they are joined by a branch of unstable solutions ---the above mentioned set of unstable equilibria (see figure \ref{fig:EtaOfDelta})---. The upper branch exists for $\delta \geq \delta_U(j)$, the lower one for $\delta \leq \delta_L(j)$. At the end points $\frac{d \eta^d}{d\delta}|_{L,U}= \infty $. In other words, solutions with large fractions of buyers, i.e. high-$\eta$ solutions, only 
exist for $\delta \ge \delta_U(j)$ whereas low-$\eta$ solutions exist only if $\delta \le \delta_L(j)$. Since $\delta_U(j) \le \delta_L(j)$, the system has multiple solutions for the demand $\eta^d$ whenever  $\delta_U(j) \le \delta \le \delta_L(j)$.

For $j=j_B$, these marginal stability curves merge at a single (degenerate) point $\delta_L(j_B) = \delta_U(j_B) = \delta_B$ with
\begin{equation}
\delta_B \equiv - \Gamma'(\eta_B) \; \eta_B + \Gamma(\eta_B)
\label{eq:deltaB}
\end{equation}
This defines the bifurcation point $B$ in the $(j, \delta)$ plane,
\begin{equation}
B \equiv \{j_B, \delta_B \}.
\label{eq:B}
\end{equation}
One should note that $\eta_{U,L}(j)$ and $\delta_{U,L}(j)$, which depend on $j$ (and on the function $\Gamma(.)$), are independent of $h$ and $p$. 

\subsubsection{Generic properties} 
In fact, the preceding analysis can be made more general because  
the main results may be obtained only based on the continuity and the convexity properties of $\Gamma$, 
without assuming any smoothness properties of the derivatives of~$f$. Let us consider this alternative. 

First, whatever the smoothness properties of $f$, the demand $\eta^d$ must be a decreasing function of the price: the economically acceptable values of the equilibrium  demand, $\eta^d \in [0,1]$, have to increase when $\delta$ increases ($p$ decreases). Thus, among the solutions of (\ref{eq:h-p}), the equilibria lie on the branches where ${\cal D}$ (defined by equation (\ref{eq:D})), is an increasing function of $\eta$ (for differentiable pdfs, this condition is given by equation (\ref{eq:stability_cond}) ). 

Next, let us analyze ${\cal D}(j;\eta)$ as a function of $\eta$ (see figure \ref{fig:calD}). By continuity of the function $\Gamma(\eta)$, ${\cal D}(j;\eta)$ is a continuous function of $\eta \in [0,1]$. As $\eta \rightarrow 0$, ${\cal D} \rightarrow - \infty$, and as $\eta \rightarrow 1$, ${\cal D} \rightarrow + \infty$. Since $\Gamma$ is concave on $]0,\eta_B]$, on increasing $\eta$ from $0$ within  $[0,\eta_B]$, ${\cal D}(j,\eta)$ has a maximum, $\delta_L(j)$, on this interval. $\delta_L(j)$ is by definition the Legendre transform of $\Gamma(\eta)$ restricted to $]0,\eta_B]$. For $\eta \geq \eta_B$, $\Gamma$ is convex, and thus ${\cal D}(j,\eta)$ has a minimum $\delta_U(j)$ on $[\eta_B, 1]$, which is the Legendre transform of $\Gamma(\eta)$ restricted to $[\eta_B, 1[$. Beyond this minimum, ${\cal D}(j;\eta)$ increases with $\eta$. 

Now, for $j< j_B$, 
the maximum on $[0,\eta_B]$ and the minimum on $[\eta_B, 1]$ of ${\cal D}(j;\eta)$ are both reached at $\eta_B$, hence
${\cal D}(j;\eta)$ increases monotonically as a function of $\eta \in [0, 1]$.
Therefore, the solutions $\eta^d$ to equation (\ref{eq:h-p}) are unique monotonically increasing functions of $\delta$ for each $j$. As a result, the inverse demand (\ref{eq:p^d}) is a uniquely defined continuously decreasing function of $\eta \in [0,1]$. 

For $j> j_B$, the maximum $\delta_L(j)$ is reached at $\eta=\eta_L(j) \in ]0,\eta_B[$. Beyond this maximum, ${\cal D}(j;\eta)$ decreases as $\eta$ increases. The minimum $\delta_U(j)$ is reached at $\eta=\eta_U(j) \in ]\eta_B, 1[$: there is an intermediate interval $]\eta_L(j),\eta_U(j)[$ containing $\eta_B$ where ${\cal D}(j;\eta)$ decreases with $\eta$, from $\delta_L$ to $\delta_U$. No value of $\eta$ within this interval can be a stable economic equilibrium. Hence, for $\delta$ ranging between these extrema of ${\cal D}(j;\eta)$ the demand $\eta^d(j;\delta)$ as a function of $\delta$ has two branches, a lower one for $\delta \leq \delta_L$, with $\eta_L^d(j;\delta) \leq \eta_L(j) < \eta_B$ and and an upper one for $\delta \geq \delta_U$, with $\eta_U^d(j;\delta) \geq \eta_U(j) > \eta_B$. 

In the case of a continuously differentiable function, the preceding results are recovered, since the Legendre transforms ---the above mentioned minimum and maximum
of ${\cal D}(j;\eta)$ for $j> j_B$--- are reached at the values of $\eta$ solutions of (\ref{eq:jdmargin=0}).
All this discussion based on convexity arguments can be extended to multimodal pdfs, that is to cases where the 
distributions of the IWPs have more than one maximum. This is done in Appendix \ref{sec:app_multimodal}.

\subsubsection{The phase diagram}
\label{sec:demand_phasediag}

\begin{figure}
\centering
\includegraphics[width=0.7\textwidth]{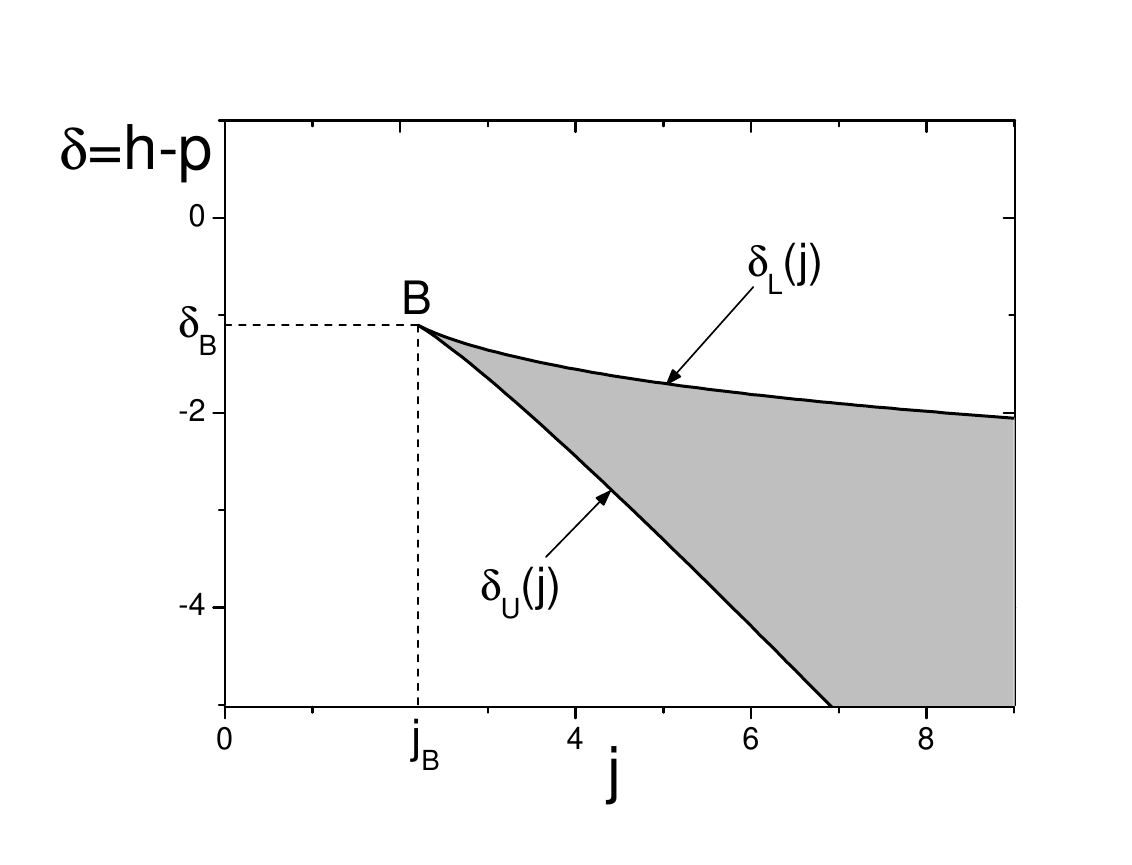}
\caption[]{\em Demand phase diagram on the plane $(j=J/\sigma, \; \delta=(H-P)/\sigma)$, for a smooth IWP distribution (here the logistic). In the shaded region the demand presents multiple Nash equilibria. Outside this region, the demand is a single valued function of $j$ and $\delta$.}
\label{fig:demandphasediagram}
\end{figure}

\begin{figure}
\centering
\includegraphics[width=0.7\textwidth]{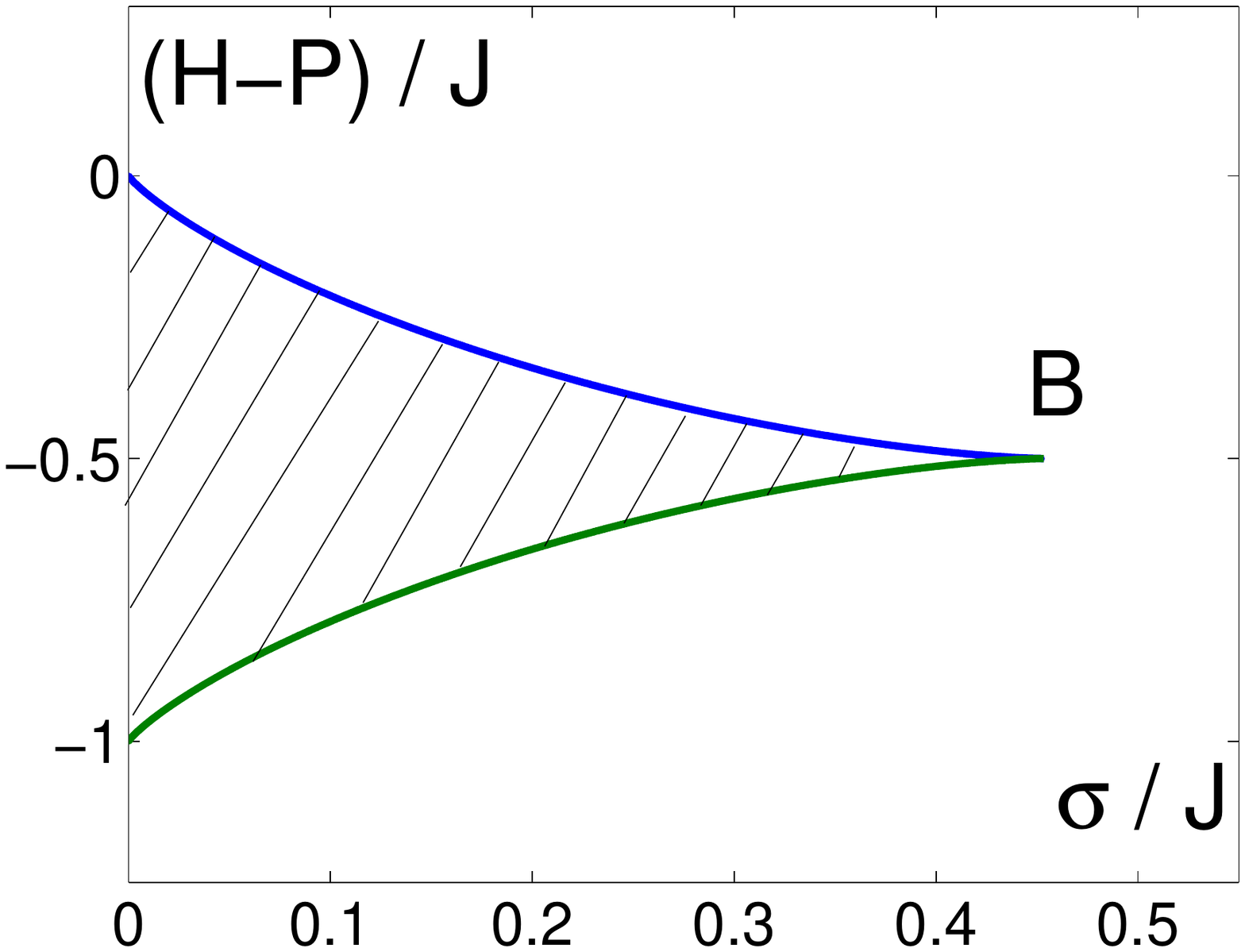}
\caption[]{\em Demand phase diagram in the plane $(\tilde \sigma= \sigma/J, \; \tilde \delta=(H-P)/J)$, for a smooth IWP distribution (here the logistic). Inside the dashed region the demand presents multiple Nash equilibria. Outside this region, the demand is a single valued function of $\tilde \sigma$ and $\tilde \delta$.}
\label{fig:demandphasediagram_deltasigma}
\end{figure}

The results of the preceding section may be summarized on a {\em customers phase diagram} in the plane $(j, \delta)$, where we represent the boundary of the multiple solutions region, as in figure \ref{fig:demandphasediagram}. These boundaries are the functions $\delta_\Lambda(j)$, ($\Lambda= L,U$) defined by equations (\ref{eq:delta1delta2}), which are the two branches of the Legendre transform of $\Gamma(\eta)$, one for $\eta < \eta_B$ and the other for $\eta > \eta_B$. 
Note that in term of prices, the extreme values $\delta_L(j)$ and $\delta_U(j)$ correspond to prices $p_L(j,h) < p_U(j,h)$ given by 
\begin{equation}
p_\Lambda(j,h) = h - \delta_\Lambda(j), \;\;\Lambda=U,L.
\label{eq:pk}
\end{equation}

By construction of the Legendre transforms, the branch $\delta=\delta_{U}(j)$ is concave, and the branch $\delta=\delta_{L}(j)$ is convex. In addition, under the smoothness hypothesis, along each branch of the multiple solutions region in the phase diagram:
\begin{equation}
\frac{ d \delta_\Lambda(j) }{dj } = \frac{d  {\cal D}(j,\eta_\Lambda(j))}{d j}=-\eta_\Lambda(j),  \;\;\; ;  \;\;\;  \Lambda \in \{L,U\}.
\label{eq:slope}
\end{equation}
This property may be easily checked by deriving (\ref{eq:delta1delta2}) with respect to $j$ and making use of (\ref{eq:jdmargin}). 
This means that the tangents to the boundaries have a slope given by the value of $\eta$ 
that is marginally stable on the corresponding boundary 
(i.e. by the $\eta$ value of the solution which appears/disappears as one crosses the boundary). 
Consequently, in the phase diagram, the width along the $\delta$-axis 
of the multiple solutions region increases with $j$ as a result of the convexity properties of the functions $\delta_\Lambda(j)$, ($\Lambda= L,U$). This may also be seen from (\ref{eq:eta12}), since the slope of the $L$ boundary (the one corresponding to $\eta_L$) is larger than that of the $U$ boundary (defined through $\eta_U$). At the bifurcation point $B$, these two boundaries merge, and, according to~(\ref{eq:slope}), have a common slope $-\eta_B$. 

Referring back to figure \ref{fig:calD}, upon increasing $\delta$ from $-\infty$, we have the following picture: if $j<j_B$, the fraction $\eta$ increases smoothly from $0$ and reaches its upper value $1$ for $\delta \rightarrow \infty$. That is, to each value of the bare surplus $\delta$, ---or each value of the average willingness to adopt, in non-market situations--- corresponds a unique fraction of buyers/adopters. In the phase diagram, figure \ref{fig:demandphasediagram}, these solutions lie on the white region. On the other hand, if $j>j_B$, when $\delta$
reaches the value $\delta_U(j)$, a second, high-$\eta$ solution appears besides the low-$\eta$ one. These solutions co-exist for $\delta_U(j) \le \delta \le \delta_L(j)$. The low-$\eta$ solution disappears when $\delta$ increases beyond $\delta_L(j)$, leaving only the high-$\eta$ solution. The parameter values for which there are multiple equilibria is the grey region of the phase diagram, figure \ref{fig:demandphasediagram}. Notice that in this region, there exists a third solution that we neglected because it corresponds to the unstable situation where the demand would increase with the price (or decrease with the bare surplus). 

As mentioned in Section \ref{sec:basic_equations}, it is also useful to consider the same results in terms of the
parameters $\tilde \sigma= \sigma/J$ and $\tilde \delta=(H-P)/J$ (see (\ref{eq:reduced_var_bis}). The phase diagram in the plane 
$(\tilde \sigma, \tilde \delta)$ is shown on figure \ref{fig:demandphasediagram_deltasigma}.
For large heterogeneity ($\sigma/J$ larger than ${\tilde \sigma}_B \equiv 1/j_B$), there is a single 
smooth solution. For weak enough heterogeneity (${\tilde \sigma}_B < 1/j_B$), 
there is a domain with multiple solutions. In the limit $\tilde \sigma \rightarrow 0$, one recovers the simple
results for a homogeneous population, as briefly discussed Section \ref{sec:hints}.

\subsubsection{Vicinity of the bifurcation point} 
\label{sec:demand_B}

The vicinity of the bifurcation point $B$ in the phase diagram is of particular interest. Under the smoothness assumption H2, 
we can study analytically its properties. Let us consider $j= j_B + \epsilon$ with $0< \epsilon << 1$. 
Expanding (\ref{eq:jdmargin}) about $j_B$, remembering that $\Gamma''(\eta_B)=0$, one gets, 
to the lowest order in $\epsilon$, the singular behavior
\begin{subequations} \label{eq:nearB}
\begin{gather}
\eta^d_{L,U} = \eta_B \pm \sqrt{\frac{2}{\Gamma'''(\eta_B)}} \; \epsilon^{1/2},  \label{eq:nearB_eta}  \\
\delta_{L,U}(j) = \Gamma(\eta_B) -\eta_B \; j_B - \eta_B \epsilon \mp \frac{2}{3} \sqrt{\frac{2}{\Gamma'''(\eta_B)}} \; \epsilon^{3/2}.  \label{eq:nearB_delta}
\end{gather}
\end{subequations}
 
The above singular behaviors are typical examples of scaling properties which are {\em universal}: the
same scaling is obtained for any smooth distribution. From
studies in statistical physics one expects the exponents (e.g., here, $1/2$ for the behavior of
$\eta$) to depend mainly on the structure of the network of interactions: the exponents
would be different at the analogous critical point for the model with agents situated on the
vertices of a $d$-dimensional square lattice and interacting only with their nearest neighbors. 
Typically the exponents depend on $d$ up to some critical dimension $d_c$, above which
they become identical to the ``mean-field" exponents, which are those obtained here with the global neighborhood.
For the present model, other universal scaling properties have been obtained, in relation with the hysteresis effects \cite{Sethna93},
and these have been used in order to analyze empirical socio-economics data \cite{MichardBouchaud}.
In a related work (with a generalization of the model to more than two choices), Borghesi
and Bouchaud \cite{BorghesiBouchaud} analyze empirical data for which the social
strength can be estimated, and is found to be close to the critical value (the
analogous of $j_B$).

In \cite{mono}, where we consider the 
supply side, it will be seen that
the bifurcation point $B=\{j_B, \delta_B \}$ in the $(j,\delta)$ plane gives 
a singular point $\{j_B, h_B \equiv \delta_B \}$ in the $(j,h)$ plane
which plays an important role in the phase diagram associated
to the optimal strategy for the monopolist.

\subsubsection{Pareto optimality and coordination}
\label{sec:Pareto}
Each one of the equilibria $\eta^d(j;\delta)$ discussed in the preceding section is a Nash equilibrium for the customers, at a posted price $p$. In this section we show that, whenever multiple solutions exist, that is for $j>j_B$, the solution with the largest $\eta$ is Pareto optimal. This is the solution $\eta^d(j;\delta)$ that satisfies $\eta^d(j;\delta) \geq \eta_U(j)$.

Let us recall that if a customer $i$ decides to buy, it is because his (normalized) surplus
\begin{equation}
s_i=\delta+j\eta+x_i,
\label{eq:s_i2}
\end{equation}
is positive. When $s_i<0$ he doesn't buy and his surplus vanishes. Thus, his actual surplus is $w_i=s_i \omega_i$ (see Section \ref{sec:model}). 

Consider now the two equilibria $\eta^d(j;\delta)$ in the region $\delta_U(j) < \delta < \delta_L(j)$ 
(see the curve ${\cal D}(j;\eta)$ for $j=5$ on figure \ref{fig:calD}).
Let's denote by $\eta_L^d(j;\delta)$ the low-$\eta$ equilibrium ($\eta_L^d(j;\delta)\leq \eta_L(j)$),
and by $\eta_U^d(j;\delta)$ the high-$\eta$ equilibrium ($\eta_U(j) \leq \eta_U^d(j;\delta)$).
In either equilibrium, the agents who buy are those with $x_i > - \delta - j \eta^d(j;\delta)$. Since $\eta^d_L(j;\delta) <  \eta^d_U(j;\delta)$, agents with $x_i < - \delta - j \eta^d_U(j)$ are not buyers in any of the equilibria whereas agents with $x_i > - \delta - j \eta^d_L(j)$ are buyers in both equilibria. Those with $- \delta - j \eta^d_U(j) < x_i < - \delta - j \eta^d_L(j)$ are buyers only in the high-$\eta$ equilibrium, and their utility is thus larger (strictly positive instead of zero) in that case. Moreover, even those agents that would buy in both cases have a larger surplus if the realized equilibrium is the high-$\eta$ one. Hence, in the high-$\eta$ equilibrium all these agents have a larger surplus than in the low-$\eta$ one. 
This situation with two possible Nash equilibria, where the strictly
dominant one may be risk dominated, is reminiscent of coordination
problems in game theory. For a detailed analysis of this analogy
see Phan and Semeshenko \cite{PhaSe07}. The present analysis shows
that coordination problems may arise in systems with heterogeneous
agents whenever the externalities are strong enough. 

Whether a Nash equilibrium ---and which one in the case of multiple equilibria--- is actually realized depends on the rationality of the agents and the information they have access to. In the context of bounded rationality and of repeated choices, a natural hypothesis is that agents estimate what will be the  fraction of adopters, and may base their estimate on previous observations. In this paper we will not discuss these issues, that we are currently analyzing. Some partial results (dynamics with myopic agents and with various reinforcement learning paradigms) are discussed elsewhere \cite{GoNaPhVa05,SemGorNadPhan07}.

\subsection{Summary of the generic customers' model}
\label{sec:demand-summary}

\begin{figure}
\centering
\includegraphics[width=0.5\textwidth]{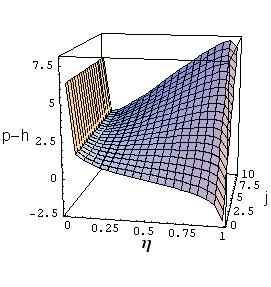}
\caption[]{\em Inverse demand $p-h\;$ ($=-\delta$) as a function of $\eta$ for different externality strength values $j$, illustrated on the case of a logistic IWP distribution.}
\label{fig:-delta(eta,j)}
\end{figure}

To summarize this section, if the social influence is small enough to satisfy the condition $j<j_B$, at each value of 
the bare surplus $\delta=h-p$, which measures the gap between the population average willingness to pay and the price, there is a unique solution $\eta^d(j;\delta)$ to equation (\ref{eq:h-p}). This demand is a monotonic increasing function of $\delta$. However, if the social influence is large enough ($j> j_B$), there is a range of values $\delta_U(j) \leq \delta \leq \delta_L(j)$ for which two different (stable) solutions exist, a high demand one ($\eta^d(j;\delta) \geq \eta_U(j) > \eta_B$) and a low demand one ($\eta^d(j;\delta) \leq \eta_L(j) < \eta_B$). In this region, the customers are faced with a coordination problem. If $\delta$ is modified dynamically within this range, the demand may jump abruptly between these two solutions, a situation analogous to so called {\em first order phase transitions} in physics. Outside the range $[\delta_U(j),\delta_L(j)]$, there is a single solution, like in the small $j$ case. 

We showed that the threshold $j_B$, which corresponds to the onset of a bifurcation in the customers 
phase diagram, is determined by the maximum $f_B$ of the IWP pdf: $j_B=1/f_B$. Although most of the analysis has been done 
for smooth pdfs, we have shown that the generic behavior stems only from convexity properties of the function $\Gamma(\eta)$, that is from the fact that the pdf $f(x)$ is strictly increasing for $x<x_B$ and strictly decreasing for $x>x_B$, 
where $x_B$ is the mode of $f$.

Some specific properties which arise for distributions with compact support are discussed 
in appendix \ref{sec:compactsupport}, and the case of distributions with infinite variance 
is studied in appendix \ref{sec:fat}. In appendix \ref{sec:app_multimodal} we extend the 
results of this section to multimodal distributions.

In market contexts (and in particular for the 
market analysis of \cite{mono}), it is useful to consider the inverse demand $p^d$ instead of $\delta$, as in standard approaches. In figure \ref{fig:-delta(eta,j)} we plot the values of $\delta = p-h$ as a function of the demand $\eta$ and the strength $j$ of the social externality for the case of a logistic distribution.

\section{Conclusion and perspectives}
\label{sec:conclusion}

The model of collective behavior considered in this paper, under the general hypothesis detailed in Section \ref{sec:model}, may be declined in both non-economic and economic contexts. In the first case, one is interested in the fraction of adopters, which corresponds to studying the demand function for an exogenous price in the second case.

Like in many other models in the recent literature, we consider optimizing agents making binary choices, with willingnesses to adopt that depend {\em additively} on an idiosyncratic part (IWP) and on the choices of other agents. The population is intrinsically heterogeneous: the IWPs are drawn from some distribution of mean $H$ and variance $\sigma^2$. In contrast with other most studied models, here the idiosyncratic willingness-to-adopt heterogeneity is frozen: it does not result from (time varying) random shocks. In other words, each agent's choice is deterministic with a well known (to him) IWP, and we concentrate on the aggregate behaviors. We analyze the equilibrium properties (Nash equilibria) characterized by the emergence of a collective behavior resulting from the combined effect of externalities and heterogeneity. 

Our results, for global uniform interactions ---a global social influence of uniform strength--- in the limit of an infinite population ---through the application of the central limit theorem--- are summarized on {\em phase diagrams}. 
The axes of a phase diagram are the relevant model parameters. 
In the case of the demand, the parameters are: $\delta$, the bare surplus (that is, in the economic context, the gap between the average IWP and the posted price), and $j$, the social influence strength, both parameters being measured in units of the standard deviation $\sigma$ of the idiosyncratic term distribution (see \ref{sec:basic_equations}).
In this space of parameters, one draws
the boundaries between regions (``phases'') of qualitatively different collective behaviors. The boundaries,
where ``phase transitions'' occur, are lines of non-analyticity (e.g. the demand is discontinuous on the boundary).
In our problem, the main feature characterizing a given region is the number of equilibria in this region.

From a constructivist point of view, our model encompasses the classical downward sloping demand curve as a particular case.
Indeed, one of the main results for the demand is that, for very general IWP distributions, there is a region in the phase diagram with multiple equilibria. More precisely, if the IWP distribution is mono-modal, there are two Nash equilibria for any $j$ larger than a distribution-dependent value $j_B$. For smaller values of $j$, the (Marshallian) demand curves are, {\it ceteris paribus}, downward sloping (i.e. monotonically decreasing with increasing prices). For large externality strengths ($j>j_B$), when the population average willingness to pay is small enough, the demand becomes not-monotonic (as in Becker's example \cite{Becker91}). This is a very general property of the model with additive externalities, and does not depend on the particular statistical distribution of the idiosyncratic preferences. 
We also discuss (although more briefly) the results for multimodal distributions 
---for which there exist several regions with multiple equilibria, with possibly more than two equlibria for some of them---, and present detailed analysis of many illustrative examples.

An important contribution of this paper is to exhibit the detailed properties of the boundaries of the regions (in the parameters space) where multiple solutions exist.
These properties are generic, in that they depend only on qualitative features of the IWP distribution.

Future work may extend the results presented in this paper in several directions. First, the individual preferences may include a stochastic (noise) term like in \cite{Fol74,Durlauf01,WeisbuchKirman,Brock93,Blume93,Blume95,NaWeChKi,ArDuLa}, on top of the idiosyncratic one. Second, the present paper concentrates on the equilibrium properties -- that may be considered as the ``static'' analysis of long term equilibria in the Marshallian tradition. Further studies should focus on the process that makes the system reach one or the other of the possible equilibria. A first study, implying revision of
beliefs in an repeated choice setting, has already been published \cite{SemGorNadPhan07}: in the region with multiple equilibria, 
interesting complex dynamics occur with a large family of different equilibria being reached, depending on the particular learning rule
used by the agents.
Third, literature on marketing and studies of social psychology shows that choices very often depend on imitation effects or social influence. For example, the existence of externalities in the Communication and Information Technologies (CIT) sector is well established, and may result in a multiplicity of equilibria \cite{Roh01}. This may arise in other sectors also. Yet, empirical and econometric studies allowing identification of the corresponding preferences distributions and the strength of the social influence are lacking \cite{BlumeDurlauf}. 
Fourth, the influence of social networks topologies deserves further attention. Ioannides \cite{Ioa06} reported results for the ``Thurnstone" model, i.e. homogeneous IWP and stochastic (logistic) choices, mainly for tree-like and one-dimensional networks with nearest-neighbor interactions. It would be interesting to explore how the phase diagrams for the model considered here ---heterogeneous IWPs and deterministic choices--- are affected by short range interactions. Analytical and simulation studies on the RFIM \cite{Sethna93,PhaPajNad03} show that the heterogeneity introduces hysteretic effects in the dynamics, with interesting path-dependant properties (return-point memory effect  \cite{Sethna93}). A statistical method to calculate the return points exactly, starting from an arbitrary initial state, has been recently proposed \cite{Shu00} for the simple case of a one dimensional periodic network with nearest-neighbor interactions (called cyclic topology in  \cite{Ioa06}). The impact of such properties on both individual and collective economic behavior remains to be investigated.

In a companion paper  \cite{mono}, we will discuss within the market context 
the profit optimization by a monopolist fixing the price.
Other extensions of the results of this second part involve the study of how, with repeated choices, the long term equilibria depend on the entangled dynamics where customers and monopolist learn from each other. Finally, at least two directions are worth to be explored: the case of an oligopolistic competition and the consequences of Coase conjecture in the case of choices with externalities involving a durable good -- an issue already addressed in the literature \cite{Manson2000}, but
not yet in the regime where multiple equilibria exist.

\subsubsection*{Acknowledgements}
This work is part of the project ``ELICCIR" supported by
the joint program ``Complex Systems in Human and Social Sciences" of the French Ministry of Research
and of the CNRS. M.B. G., J.-P. N. and D. P. are CNRS members. 
This work has been partly done
while M.B. G. and V. S. were with the laboratory LEIBNIZ, CNRS-IMAG, Grenoble.

\newpage

\bibliographystyle{plain}
\bibliography{cust_bib}

\begin{thebibliography}{10}

\bibitem{AnDeTh}
S.~P. Anderson, A.~de~Palma, and J.-F. Thisse.
\newblock {\em Discrete Choice Theory of Product Differentiation}.
\newblock MIT Press, Cambridge MA, (1992).

\bibitem{Arthur94}
W.~B. Arthur.
\newblock El farol.
\newblock {\em Amer. Econ. Review}, {\bf 84}:406--, (1994).

\bibitem{Arthur99}
W.~B. Arthur.
\newblock Complexity and the economy.
\newblock {\em Science}, {\bf 284}:107--109, (1999).

\bibitem{ArDuLa}
W.B. Arthur, S.N. Durlauf, and D.A.~Lane Eds.
\newblock {\em The Economy as an Evolving Complex System II}.
\newblock Santa Fe Institute Studies in the Sciences of Complexity, Volume
  XXVII, Addison-Wesley Pub.Co., Reading Ma., (1997).

\bibitem{Art73}
R.~Artle and C.~Averous.
\newblock The telephone system as a public good: Static and dynamic aspects.
\newblock {\em The Bell Journal of Economics and Management Science}, {\bf 4}
  (1):89--100, (1973).

\bibitem{BaldwinKrugman}
R.~Baldwin and P.~Krugman.
\newblock Persistent trade effects of large exchange rate shocks.
\newblock {\em Quarterly Journal of Economics}, {\bf 419}:635--654, (1989).

\bibitem{Ball}
P.~Ball.
\newblock The physical modelling of human social systems.
\newblock {\em Complexus}, {\bf 1}:190--206, (2003).

\bibitem{Bec74}
G.~S. Becker.
\newblock A theory of social interactions.
\newblock {\em The Journal of Political Economy}, {\bf 82} (6):1063--1093,
  (1974).

\bibitem{Becker91}
G.~S. Becker.
\newblock A note on restaurant pricing and other examples of social influences
  on price.
\newblock {\em The Journal of Political Economy}, {\bf 99}:1109--1116, (1991).

\bibitem{Bec00}
G.~S. Becker and K.M Murphy.
\newblock {\em Social Economics. Market Behavior in a Social Environment}.
\newblock Cambrige Ma. The Belknap Press \& Harvard University Press, (2000).

\bibitem{Ben96}
R.~Benabou.
\newblock Equity and efficiency in human capital investment: The local
  connection.
\newblock {\em Review of Economic Studies}, {\bf 63}:237--264, (1996).

\bibitem{Ber94}
B.D. Bernheim.
\newblock A theory of conformity.
\newblock {\em Journal of Political Economy}, {\bf 102}:841--77, (1994).

\bibitem{Blume93}
L.~E. Blume.
\newblock The statistical mechanics of strategic interaction.
\newblock {\em Games and Economic Behavior}, {\bf 5}:387--424, (1993).

\bibitem{Blume95}
L.~E. Blume.
\newblock The statistical mechanics of best-response strategy revisions.
\newblock {\em Games and Economic Behavior}, {\bf 11}:111--145, (1995).

\bibitem{BlumeDurlauf}
L.~E. Blume and S.~N. Durlauf.
\newblock Identifying social interactions: a review.
\newblock {\em Working paper}, (2005).

\bibitem{BorghesiBouchaud}
C.~Borghesi and J.-P. Bouchaud.
\newblock Of songs and men: a model for multiple choice with herding.
\newblock {\em Working paper http://fr.arxiv.org/abs/physics/0606224}, 2006.

\bibitem{BouchaudPotters}
J.-P. Bouchaud and M.~Potters.
\newblock {\em Theory of Financial Risk and Derivative Pricing}.
\newblock Cambridge University Press, 2nd Edition, 2003.

\bibitem{Bouchaud}
J-Ph. Bouchaud.
\newblock Power-laws in economics and finance: some ideas from physics.
\newblock {\em Quantitative finance}, {\bf 1}:105, (2000).

\bibitem{Brock93}
W.~A. Brock.
\newblock Pathways to randomness in the economy: Emergent non-linerarity and
  chaos in economic and finance.
\newblock {\em Estudios Economicos}, {\bf 8}:1:3--55, (1993).

\bibitem{BrockDurlauf01a}
W.~A. Brock and S.~N. Durlauf.
\newblock Discrete choice with social interactions.
\newblock {\em Review of Economic Studies}, {\bf 68}:235--260, (2001).

\bibitem{BrockDurlauf01b}
W.~A. Brock and S.~N. Durlauf.
\newblock Interactions-based models.
\newblock In Heckman J. and Leamer E., editors, {\em Handbook of Economics},
  volume~{\bf 5}, Amsterdam, (2001). North-Holland.

\bibitem{BuGeKl}
J.~I. Bulow, J.~D. Geanalopolos, and P.~D. Klemperer.
\newblock Multimarket oligopoly: strategic substitutes and complements.
\newblock {\em Journal of Political Economy}, {\bf 93/3}:488--511, (1985).

\bibitem{Cooper99}
R.~W. Cooper.
\newblock {\em Coordination Games, Complementarities and Macroeconomics}.
\newblock Cambridge University Press, (1999).

\bibitem{Crane91}
J.~Crane.
\newblock The epidemic theory of ghettos and neighborhood effects of dropping
  out and teenage childbearing.
\newblock {\em American Journal of Sociology}, {\bf 96}:1226--1259, (1991).

\bibitem{Cur87}
N.~Curien and M.~Gensollen.
\newblock Les th\'eories de la demande de raccordement t\'el\'ephonique.
\newblock {\em Revue Economique}, {2 mars}:203--255, 1987.

\bibitem{Deffuant01}
G.~Deffuant, D.~Neau, F.~Amblard, and G.~Weisbuch.
\newblock Mixing beliefs among interacting agents.
\newblock {\em Advances in Complex Systems}, {\bf 3}:87--98, (2001).

\bibitem{Durlauf91}
S.~N. Durlauf.
\newblock Multiple equilibria and persistence in aggregate fluctuations.
\newblock {\em American Economic Review}, {\bf 81}:70--74, (1993).

\bibitem{Durlauf93}
S.~N. Durlauf.
\newblock Nonergodic economic growth.
\newblock {\em Review of Economic Studies}, {\bf 60}:203:349--366, (1993).

\bibitem{Durlauf94}
S.~N. Durlauf.
\newblock Path dependance in aggegate output.
\newblock {\em Industrial and Corporate Change}, {\bf 1}:149--172, (1994).

\bibitem{Durlauf96}
S.~N. Durlauf.
\newblock A theory of persistent income inequality.
\newblock {\em Journal of Economic Growth}, {\bf 1}:349--366, (1996).

\bibitem{Durlauf97}
S.~N. Durlauf.
\newblock Statistical mechanics approaches to socioeconomic behavior.
\newblock In B.~Arthur, S.~N. Durlauf, and D.~Lane, editors, {\em The Economy
  as an Evolving Complex System II}. Santa Fe Institute Studies in the Sciences
  of Complexity, Volume XVII, Addison-Wesley Pub. Co, (1997).

\bibitem{Durlauf99}
S.~N. Durlauf.
\newblock How can statistical mechanics contribute to social science?
\newblock {\em Proceedings of the National Academy of Sciences}, {\bf
  96}:10582--10584, (1999).

\bibitem{Durlauf01}
S.~N. Durlauf.
\newblock A framework for the study of individual behaviour and social
  interactions.
\newblock {\em Working paper}, {}, (2001).

\bibitem{Durlauf06}
S.~N. Durlauf.
\newblock Groups, social influences, and inequality: A membership theory
  perspective on poverty traps.
\newblock In S.~Bowles, S.~N. Durlauf, and K.~Hoff, editors, {\em Poverty
  Traps}. Princeton: Princeton University Press, (2006).

\bibitem{EhMaVe04}
G.~Ehrhardt, M.~Marsili, and F.~Vega-Redondo.
\newblock Emergence and resilience of social networks: a general theoretical
  framework.
\newblock {\em Working paper, ArXiv e-prints,
  http://arxiv.org/abs/physics/0504124}, (2005).

\bibitem{Fol74}
H.~F{\"{o}}llmer.
\newblock Random economies with many interacting agents.
\newblock {\em Journal of Mathematical Economics}, {\bf 1}:1:51--62, (1974).

\bibitem{Galam82}
S.~Galam.
\newblock A new multicritical point in anisotropic magnets. {III}.
  {F}erromagnets in both a random and a uniform longitudinal field.
\newblock {\em Journal of Physics C}, {\bf 15}:529--545, (1982).

\bibitem{GalAha80}
S.~Galam and A.~Aharony.
\newblock New multicritical points in anisotropic magnets, i. ferromagnets in a
  random longitudinal field.
\newblock {\em Journal of Physics C}, {\bf 13}:1065--1081, (1980).

\bibitem{GaGeSh}
S.~Galam, Y.~Gefen, and Y.~Shapir.
\newblock Sociophysics: A mean behavior model for the process of strike.
\newblock {\em Mathematical Journal of Sociology}, {\bf 9}:1--13, (1982).

\bibitem{GlaeserScheinkman}
E.~Glaeser and J.~A. Scheinkman.
\newblock Non-market interactions.
\newblock In M.~Dewatripont, L.P. Hansen, and S.~Turnovsky, editors, {\em
  Advances in Economics and Econometrics: Theory and Applications, Eight World
  Congress}. Cambridge University Press, (2002).

\bibitem{GKSS92}
E.~L. Glaeser, H.~D. Kallai, J.~A. Scheinkman, and A.~Shleifer.
\newblock Growth in cities.
\newblock {\em Journal of Political Economy}, {\bf 100}:6:1126--1152, (1992).

\bibitem{GlaeSaceSche}
E.~L. Glaeser, B.~Sacerdote, and J.~A. Scheinkman.
\newblock Crime and social interactions.
\newblock {\em Quarterly Journal of Economics}, {\bf CXI}:507--548, (1996).

\bibitem{Gordon04}
M.~B. Gordon.
\newblock An introduction to statistical mechanics.
\newblock In Bourgine P. and Nadal J-P., editors, {\em Cognitive Economics},
  pages 131--155. Springer, (2004).

\bibitem{mono}
M.~B. Gordon, J.-P. Nadal, D.~Phan, and V.~Semeshenko.
\newblock The perplex monopolist: Optimal pricing with customers under social
  influence.
\newblock {\em In preparation}, (2007).

\bibitem{GoNaPhVa05}
M.~B. Gordon, J.-P. Nadal, D.~Phan, and J.~Vannimeuns.
\newblock Seller's dilemma due to social interactions between customers.
\newblock {\em Physica A}, {{\bf 356}, Issues 2-4}:628--640, (2005).

\bibitem{Granovetter78}
M.~Granovetter.
\newblock Threshold models of collective behavior.
\newblock {\em American Journal of Sociology}, {\bf 83}(6):1360--1380, (1978).

\bibitem{GranovetterSoong86}
M.~Granovetter and R.~Soong.
\newblock Threshold models of interpersonal effects in consumer demand.
\newblock {\em Journal of Economic Behavior \& Organization}, {\bf 7}
  (1):83--99, (1986).

\bibitem{Hildebrand83}
W.~Hildebrand.
\newblock On the ``laws of demand''.
\newblock {\em Econometrica}, {\bf 51}:997--1019, (1983).

\bibitem{Schweitzer01}
J.~Holyst, K.~Kacperski, and F.~Schweitzer.
\newblock Social impact models of opinion dynamics.
\newblock In D.~Stauffer, editor, {\em Annual Reviews of Computational Physics
  IX}, pages 253--273. World Scientific, Singapore, (2001).

\bibitem{Ioa06}
Y.M. Ioannides.
\newblock Topologies of social interactions.
\newblock {\em Economic Theory}, {\bf 28}:559--584, (2006).

\bibitem{Ising24}
Ernst Ising.
\newblock {\em Beitrag zur Theorie des Ferromagnetismus}.
\newblock Dissertation, (1924).

\bibitem{KAT94}
M.~L. Katz and C.~Shapiro.
\newblock Systems competition and network effects.
\newblock {\em The Journal of Economic Perspectives}, {\bf 8}(2):93--115,
  (1994).

\bibitem{KindSnell}
R.~Kindermann and J.~L. Snell.
\newblock {\em Random Markov Fields and their Applications}.
\newblock American Mathematical Society, Providence, Rhode Island, (1980).

\bibitem{Kirman92}
A.~P. Kirman.
\newblock Whom or what does the representative individual represent?
\newblock {\em Journal of Economic Perspective}, {\bf 6}(2):117--136, (1992).

\bibitem{Kra06b}
B.~Krauth.
\newblock Simulation-based estimation of peer effects.
\newblock {\em Journal of Econometrics}, forthcoming, (2006).

\bibitem{Kra06a}
B.~Krauth.
\newblock Social interactions in small groups.
\newblock {\em Canadian Journal of Economics}, {\bf 39} (2) May:414--433,
  (2006).

\bibitem{LIE50}
H.~Leibenstein.
\newblock Bandwagon, snob, and veblen effects in the theory of consumers'
  demand.
\newblock {\em Quarterly Journal of Economics}, {\bf 64}:2:183--207, (1950).

\bibitem{Luce59}
R.~D. Luce.
\newblock {\em Individual Choice Behavior}.
\newblock Wiley, New-York, (1959).

\bibitem{Manski77}
C.F. Manski.
\newblock The structure of random utility models.
\newblock {\em Theory and Decision}, {\bf 8}:229--254, (1977).

\bibitem{Man00}
C.F. Manski.
\newblock Economic analysis of social interactions.
\newblock {\em Journal of Economic Perspectives}, {\bf 14} (3) Summer:115--136,
  (2000).

\bibitem{Manson2000}
R.~Manson.
\newblock Network externalities and the coase conjecture.
\newblock {\em European Economic Review}, {\bf 44}(10):1981--1992, (2000).

\bibitem{McFadden74}
D.L. McFadden.
\newblock Conditional logit analysis of qualitative choice analysis.
\newblock In Zarembka, editor, {\em Frontiers of Econometrics}, pages 105--142.
  New York: Academic Press, (1974).

\bibitem{McFadden76}
D.L. McFadden.
\newblock Quantal choice analysis: A survey.
\newblock {\em Annals of Economic and Social Measurement}, {\bf 5} No.
  4:363--390, (1976).

\bibitem{McKelveyPalfrey95}
R.~D. McKelvey and T.~R. Palfrey.
\newblock Quantal response equilibria for normal games.
\newblock {\em Games and Economic Behavior}, {\bf 10}:6--38, (1995).

\bibitem{MichardBouchaud}
Q.~Michard and J.-P. Bouchaud.
\newblock Theory of collective opinion shifts: from smooth trends to abrupt
  swings.
\newblock {\em The European Physical Journal B - Condensed Matter and Complex
  Systems}, {\bf 47}:151--159, (2005).

\bibitem{NaPhGoVa05}
J.-P. Nadal, D.~Phan, M.~B. Gordon, and J.~Vannimenus.
\newblock Multiple equilibria in a monopoly market with heterogeneous agents
  and externalities.
\newblock {\em Quantitative Finance}, {\bf 5}(6):557--568, (2006). Presented at
  the 8th Annual Workshop on Economics with Heterogeneous Interacting Agents
  (WEHIA 2003).

\bibitem{NaWeChKi}
J.-P. Nadal, G.~Weisbuch, O.~Chenevez, and A.~Kirman.
\newblock A formal approach to market organisation: Choice functions, mean
  field approximation and maximum entropy principle.
\newblock In J.~Lesourne and A.~Orl\'ean, editors, {\em Advances in
  Self-Organization and Evolutionary Economics}, pages 149--159. Economica,
  London, (1998).

\bibitem{Onsager}
L.~Onsager.
\newblock Crystal statistics, {I}. {A} two-dimensional model with an order
  disorder transition.
\newblock {\em Physical Review}, 65:117--149, (1944).

\bibitem{Orlean90}
A.~Orl{\'e}an.
\newblock Le r\^{o}le des influences interpersonnelles dans la d\'etermination
  des cours boursiers.
\newblock {\em Revue Economique}, vol. {\bf 41}, n.5:839--868, (1990).

\bibitem{Orlean95}
A.~Orl{\'e}an.
\newblock Bayesian interactions and collective dynamics of opinion: Herd
  behaviour and mimetic contagion.
\newblock {\em Journal of Economic Behavior and Organization}, 28:257--274,
  (1995).

\bibitem{Ostrom00}
E.~Ostrom.
\newblock Collective action and the behaviour of social norms.
\newblock {\em Journal of Economic Perspectives}, {\bf 14}:137--158, (2000).

\bibitem{PhaGorNad04}
D.~Phan, M.~B. Gordon, and J.-P. Nadal.
\newblock Social interactions in economic theory: an insight from statistical
  mechanics.
\newblock In Bourgine P. and Nadal J-P., editors, {\em Cognitive Economics}.
  Springer, (2004).

\bibitem{PhaPajNad03}
D.~Phan, S.~Pajot, and J-P. Nadal.
\newblock The monopolist's market with discrete choices and network externality
  revisited: Small-worlds, phase transition and avalanches in an ace framework.
\newblock In {\em Ninth annual meeting of the Society of Computational
  Economics, University of Washington, Seattle, USA, July 11-13, 2003}, (2003).

\bibitem{PhaSe07}
D.~Phan and V.~Semeshenko.
\newblock Equilibria in models of binary choice with heterogeneous agents and
  social influence.
\newblock {\em \emph{submitted} to European Journal of Economic and Social
  Systems}, {\bf }, (2007).

\bibitem{Roh74}
J.~Rohlfs.
\newblock Theory of interdependent demand for a communications service.
\newblock {\em Bell Journal of Economics and Management Science}, Vol. {\bf 5},
  No. 1, Spring:16--37, (1974).

\bibitem{Roh01}
J.~Rohlfs.
\newblock {\em Bandwagon effects in High-Technology Industries}.
\newblock Cambridge, Ma., MIT Press, (2001).

\bibitem{Schelling73}
T.~S. Schelling.
\newblock Hockey helmets, concealed weapons, and daylight saving: A study of
  binary choices with externalities.
\newblock {\em The Journal of Conflict Resolution}, t. XVII, N. 3, (1973).

\bibitem{Schelling}
T.~S. Schelling.
\newblock {\em Micromotives and Macrobehavior}.
\newblock W.W. Norton and Co, N.LY., (1978).

\bibitem{SemGorNadPhan07}
V.~Semeshenko, M.~B. Gordon, J.-P. Nadal, and D.~Phan.
\newblock Choice under social influence: effects of learning behaviors on the
  collective dynamics.
\newblock In Topol R. and Walliser B., editors, {\em Contributions to economic
  analysis. Cognitive Economics. New Trends}, pages 177--203. Elsevier, (2007).

\bibitem{Sethna93}
J.~P. Sethna, K.~Dahmen, S.~Kartha, J.~A. Krumhansl, B.W. Roberts, and J.~D.
  Shore.
\newblock Hysteresis and hierarchies: Dynamics of disorder-driven first-order
  phase transformations.
\newblock {\em Physical Review Letters}, {\bf 70}:3347--3350, (1993).

\bibitem{ShapVar99}
C.~Shapiro and H.~Varian.
\newblock {\em Information Rules: A Strategic Guide to the Network Economy}.
\newblock Harvard Business School Press, (1999).

\bibitem{Shu00}
P.~Shukla.
\newblock Exact solution of return hysteresis loops in a one-dimensional
  random-field ising model at zero temperature.
\newblock {\em Phys. Rev. E}, {\bf 62} (4) October:4725--4729, (2000).

\bibitem{Soe06}
A.R. Soetevent and P.~Kooreman.
\newblock A discrete choice model with social interactions; with an application
  to high school teen behavior.
\newblock {\em Journal of Applied Econometrics}, forthcoming, (2006).

\bibitem{SocialPerco}
S.~Solomon, G.~Weisbuch, L.~de~Arcangelis, N.~Jan, and D.~Stauffer.
\newblock Social percolation models.
\newblock {\em Physica A}, {\bf 277}:239--247, (2000).

\bibitem{Stanley}
H.E. Stanley.
\newblock {\em Introduction to phase transitions and critical phenomena}.
\newblock Oxford University Press, (1971).

\bibitem{Con02}
G.~Topa T.~Conley.
\newblock Socio-economic distance and spatial patterns in unemployment.
\newblock {\em Journal of Applied Econometrics}, {\bf 17} (4):303--327, (2002).

\bibitem{Thurstone}
L.~L. Thurstone.
\newblock Psychological analysis.
\newblock {\em American Journal of Psychology}, {{\bf 38}}:368--398, (1927).

\bibitem{Top01}
G.~Topa.
\newblock Social interactions, local spillovers and unemployment.
\newblock {\em Review of Economic Studies}, {\bf 68} (2):261--295, (2001).

\bibitem{Valente}
T.~W. Valente.
\newblock {\em Network Models of the Diffusion of Innovations}.
\newblock Hampton Press, Cresskill, NJ, (1995).

\bibitem{vRab74}
B.~Von~Rabenau and K.~Stahl.
\newblock Dynamic aspects of public goods: a further analysis of the telephone
  system.
\newblock {\em Bell Journal of Economics and Management Science}, Vol. {\bf 5},
  n2:651--669, 1974.

\bibitem{WattsCascades}
D.~J. Watts.
\newblock A simple model of global cascades on random networks.
\newblock {\em PNAS}, Vol. {\bf 99} no. 9:5766--5771, (2002).

\bibitem{Wei00}
W.~Weidlich.
\newblock {\em Sociodynamics: A Systematic Approach to Mathematical Modelling
  in the Social Sciences}.
\newblock Harwood Academic Publishers, (2000).

\bibitem{WeidlichHaag83}
W.~Weidlich and G.~Haag.
\newblock {\em Concepts and models of a quantitative sociology}.
\newblock Berlin, Heidelberg/New York, Springer-Verlag, (1983).

\bibitem{WeisbuchKirman}
G.~Weisbuch, A.~Kirman, and D.~Herreiner.
\newblock Market organisation and trading relationships.
\newblock {\em Working paper 1996, published in: The Economic Journal}, Volume
  {\bf 110} Issue 463:411--462, (2000).

\bibitem{WeiSta}
G.~Weisbuch and D.~Stauffer.
\newblock Adjustment and social choice.
\newblock {\em Physica A}, {\bf 323}:651--662, (2003).

\bibitem{Ioa03}
J.~Zabel Y.M.~Ioannides.
\newblock Neighborhood effects and housing demand.
\newblock {\em Journal of Applied Econometrics}, {\bf 18}:563--584, (2003).

\end{thebibliography}

\newpage

\renewcommand{\theequation}{A-\arabic{equation}} 
\setcounter{equation}{0}  
\renewcommand{\thesection}{A} 
\setcounter{section}{0}  

\section{Appendix: Demand for other distributions}
\label{app:other_distr}
In this Appendix we extend our analysis to more general distributions. 

We first (\ref{sec:compactsupport}) explicit the particularities introduced on the above generic results when the pdf has a bounded support. In Section \ref{sec:fat} we relax hypothesis H5, and consider pdfs with unbounded support called {\em fat tail} distributions in the literature. Finally, we extend our results to multimodal distributions in Section \ref{sec:app_multimodal}.

\subsection{Pdfs with compact support}
\label{sec:compactsupport}
We consider here pdfs $f(x)$ with compact supports: $x \in [x_m, x_M]$ presenting a unique maximum (which may be located at one boundary). Clearly, such pdfs have finite variances. The discussion follows the same lines as that of the generic smooth distributions, except that in addition one has to pay attention to the values of $\Gamma$ and its derivatives at the boundaries $\eta=0$, $\eta=1$. A simple uniform distribution, analyzed in \cite{GoNaPhVa05}, is a particular case where the maximum of the pdf is degenerate. 

In this section, derivatives like $\Gamma'(1)$ or $\Gamma'(0)$, stand for the left and the  right derivative of $\Gamma$ at $\eta=1$ and $\eta=0$, respectively. Due to the fact that the pdf strictly vanishes beyond its support, if the price is very high with respect to $h$ (small $\delta$) there may be no buyers at all, and $\eta=0$. On the contrary, if the price is very low (large $\delta$) the market may or not saturate, i.e. $\eta=1$, depending on the behaviour of the pdf in the vicinity of $x_m$. We represent the lines delimiting the regions where these solutions exist on the phase diagram. It should be stressed that these lines only indicate saturation and non-existence of a market. Their nature is different from that of the boundaries $\delta_L$ and $\delta_U$. In the following we consider a triangular distribution with a maximum $f_B$ at $x_B$ to illustrate our general results. The figures in this section correspond to $x_B=-1$, i.e. a case where the maximum lies inside the support. The case $x_B=x_m$ has been considered in the study of the learning dynamics \cite{SemGorNadPhan07}.

\begin{figure}
\includegraphics[width=0.5\textwidth]{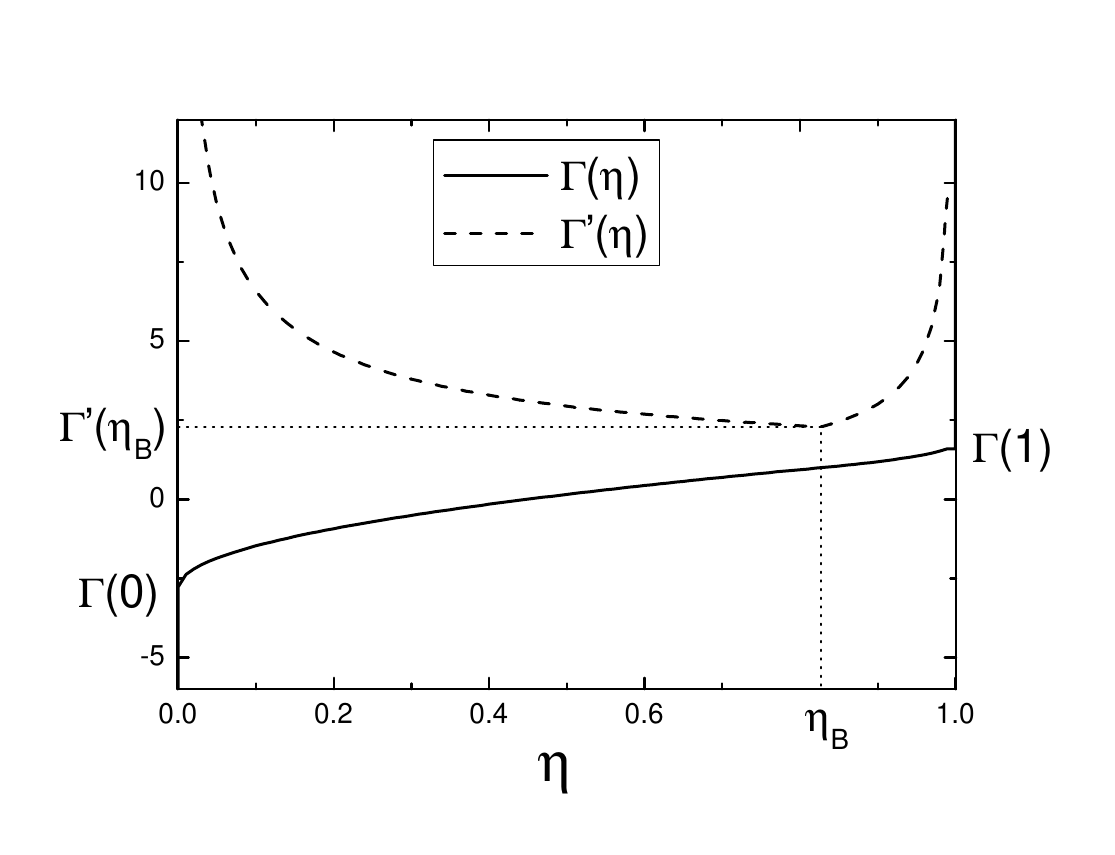}
\hfill
\includegraphics[width=0.55\textwidth]{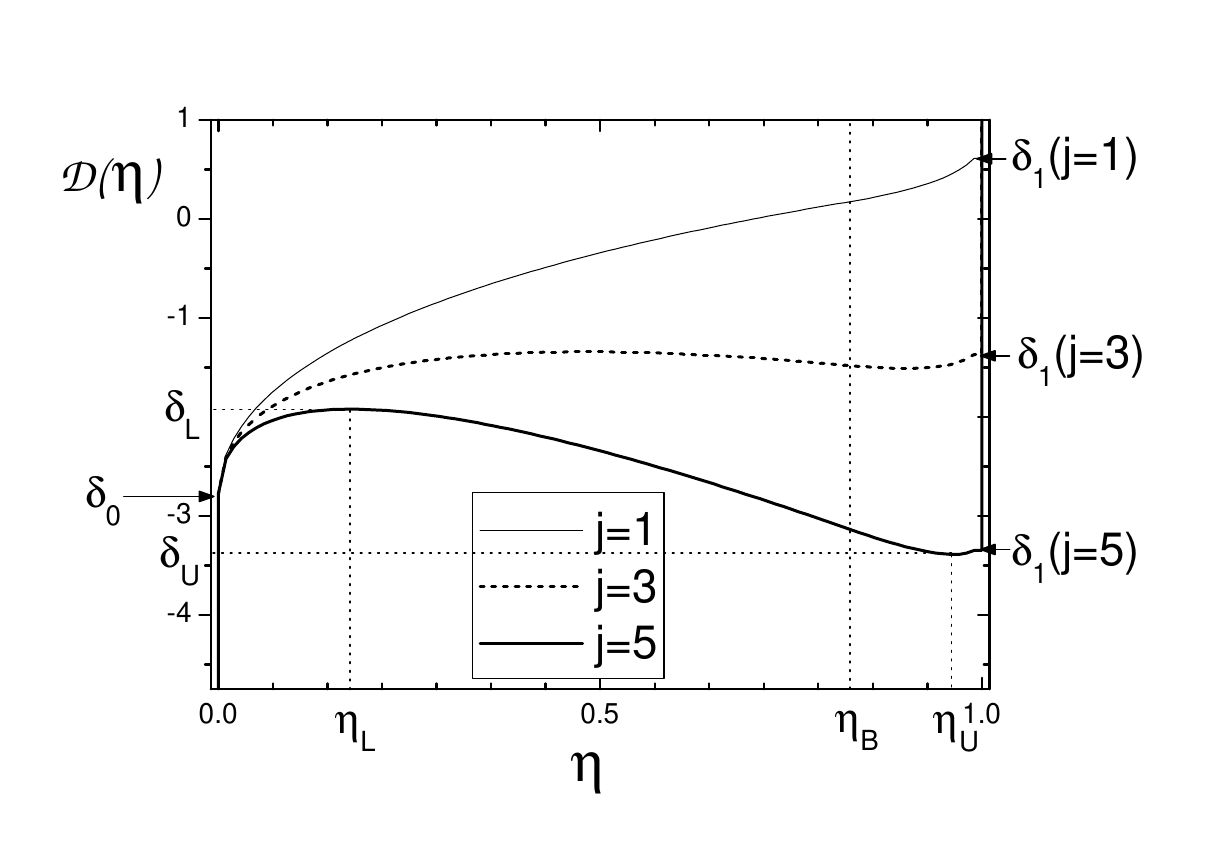}
\caption[]{{\em Triangular pdf of unitary variance and a maximum at $x_B=-1$. 
Left: $\Gamma(\eta)$ and its first derivative. 
Right: ${\cal D}(j;\eta)$ for different values of $j$. The values of $\eta_L$, $\delta_L$, $\eta_U$ and $\delta_U$
are represented for the particular value $j=5$. 
}}
\label{fig:triangM-1Gammas}
\end{figure}

In the case of compact supports $[x_m, x_M]$, $\Gamma(\eta)$ increases from $\Gamma(0)=-x_M <0$, to $\Gamma(1)= -x_m > 0$. Like in the generic case of unbounded supports, $\Gamma'$ reaches a minimum at $\eta_B$, and there is a critical value $j_B=1/f_B$ beyond which multiple solutions appear. Notice that if the maximum of the pdf lies at $x_m$ or at $x_M$, $\eta_B$ lies at one of the boundaries of the $[0,1]$ interval. As already shown in Section \ref{sec:directinverse}, if the pdf is symmetric ---as is the case for the uniform distribution---, $\eta_B=1/2$. 

In our example of a triangular pdf with maximum at $x_B$ we have:
\begin{equation}
\label{eq:triang_distribution}
f(x)=\left\{
\begin{array}{lll}
\frac{2(x-x_m)}{(x_M-x_m)(x_B-x_m)} & {\rm if} & x_m \leq x \leq x_B, \\
\frac{2(x_M-x)}{(x_M-x_m)(x_M-x_B)} & {\rm if} & x_B \leq x \leq x_M, 
\end{array}
\right.
\end{equation}
with $f(x)=0$ outside the support. At its maximum, $f_B=2/(x_M-x_m)$. The constraint of zero mean and unit variance imposes a relationship between the support boundaries and the value of $x_B$: $x_m=(-x_B-\sqrt{3(8-x_B^2)})/2$, $x_M= (-x_B+\sqrt{3(8-x_B^2)})/2$. 

If $x_B=x_M$ the distribution increases monotonically from $0$ at $x_m$, reaching its maximum at $x_M$, and is defined only by the first equation in (\ref{eq:triang_distribution}). Conversely, if $x_B=x_m$ then $f(x)$ decreases inside its support, and is defined by the second equation in (\ref{eq:triang_distribution}). In both cases, $f(x)$ presents a discontinuity at one boundary of its support.

The function $\Gamma$ and its first derivative for the triangular distribution are respectively
\begin{equation}
\label{eq:Gamma_triang_distribution}
\Gamma(\eta)=\left\{
\begin{array}{lll}
-x_M + \sqrt{(x_M-x_m)(x_M-x_B) \eta} & {\rm if} & 0 \leq \eta \leq \eta_B \\
-x_m - \sqrt{(x_M-x_m)(x_B-x_m) (1-\eta)} & {\rm if} & \eta_B \leq \eta \leq 1 
\end{array}
\right.
\end{equation}
and 
\begin{equation}
\label{eq:Gamma'_triang_distribution}
\Gamma'(\eta)=\left\{
\begin{array}{lll}
\frac{\sqrt{(x_M-x_m)(x_M-x_B)}}{2 \sqrt{\eta}} & {\rm if} & 0 \leq \eta \leq \eta_B \\
\frac{\sqrt{(x_M-x_m)(x_B-x_m)}}{2 \sqrt{(1-\eta)}} & {\rm if} & \eta_B \leq \eta \leq 1 
\end{array}
\right.
\end{equation}
where $\eta_B=(x_M-x_B)/(x_M-x_m)$ is the inflexion point of $\Gamma$. They are represented on figure \ref{fig:triangM-1Gammas} (left) for the particular value $x_B=-1$. Notice that $\Gamma''(\eta)$ is discontinuous at $\eta_B$, because the maximum of $f(x)$ is a cusp.

The corresponding inverse demand function ${\cal D}(j;\eta)$ (equation (\ref{eq:D} ) ) is represented on figure \ref{fig:triangM-1Gammas} (right). 
Due to the finite range of the compact support, there are two new particular values of $\delta$: $\delta_0 \equiv {\cal D}(j;0) = \Gamma(0)=-x_M$, independent of $j$, and $\delta_1(j) \equiv {\cal D}(j;1) = \Gamma(1) - j = -x_m-j$: for $\delta<\delta_0$, $\eta=0$ (no-market) is a solution, while for $\delta>\delta_1(j)$ there is a solution $\eta=1$ (market saturation). These extreme values of $\eta$ may be reached upon finite values of $\delta$ (i.e. finite prices and finite average IWP) only in the case of compact supports. 

For $j<j_B$, ${\cal D}(j;\eta)$ is strictly increasing and invertible on $]0,1[$: for any $\delta$ in $]-x_M, -x_m - j[$, equation (\ref{eq:h-p}) has a unique solution $\eta^d(\delta) \neq \{0,1\}$. One can easily check that $j< j_B$ implies $j \leq x_M-x_m$, so that $-x_M < -x_m - j$ and consequently  ${\cal D}(j;0) < {\cal D}(j;1)$. In the particular case where $f$ is the uniform distribution, one has precisely $j_B=x_M-x_m$. For the triangular pdf (\ref{eq:triang_distribution}), (\ref{eq:jB}) gives $j_B=(x_M-x_m)/2$, the support's half-width. Thus, for $\delta_0 < \delta < \delta_1(j)$ the fraction of buyers/adopters is a monotonic increasing function of $\delta$. For $\delta<\delta_0$, $\eta=0$, and for $\delta > \delta_1(j)$ the market saturates ($\eta=1$). 

For $j>j_B$ there are two stable solutions whenever $\delta_U(j) \leq \delta \leq \delta_L(j)$. Due to the existence of the extreme solutions $\eta=0$ and $\eta=1$, the analysis is more cumbersome than for infinite supports. If the maximum of the pdf lies inside the support, the solutions $\eta_L(j)$ and $\eta_U(j)$ of equation (\ref{eq:jdmargin=0}) lie in $]0,1[$ and $\delta_U(j)$ and $\delta_L(j)$ both satisfy ${\cal D}'=0$. On increasing $\delta$ from $-\infty$, there is no demand until $\delta=\min\{\delta_0,\delta_U(j)\}$. If $\delta_0 < \delta_U(j)$, when $\delta$ increases beyond $\delta_0$ the demand becomes finite and remains unique provided that $\delta_0 < \delta < \delta_U(j)$. For $\delta > \delta_U(j)$ we enter the region of multiple solutions. On the other hand, if $\delta_0 > \delta_U(j)$, the system steps directly from the no-demand solution into a region where a finite demand equilibrium with $\eta > \eta_B$ coexists with the no-demand one. 
Notice that the high-$\eta$ solution may correspond to either a fraction of buyers strictly smaller than $1$ (if $\delta_L(j) < \delta_1(j)$) or to saturation (if $\delta_L(j) > \delta_1(j)$). 
In the case of the triangular distribution it is straightforward to check that the multiple solutions region sets in at $j_B=(x_M-x_m)/2$, $\delta_B=x_B/2$.

If the pdf has its maximum at one of the boundaries of its support, either $\eta_U$ or $\eta_L$ coincide with $\eta_B$. More precisely, if $x_B=x_m$ then $\eta_U=\eta_B=1$, if $x_B=x_M$, $\eta_L=\eta_B=0$.

\begin{figure}
\centering
\includegraphics[width=0.8\textwidth]{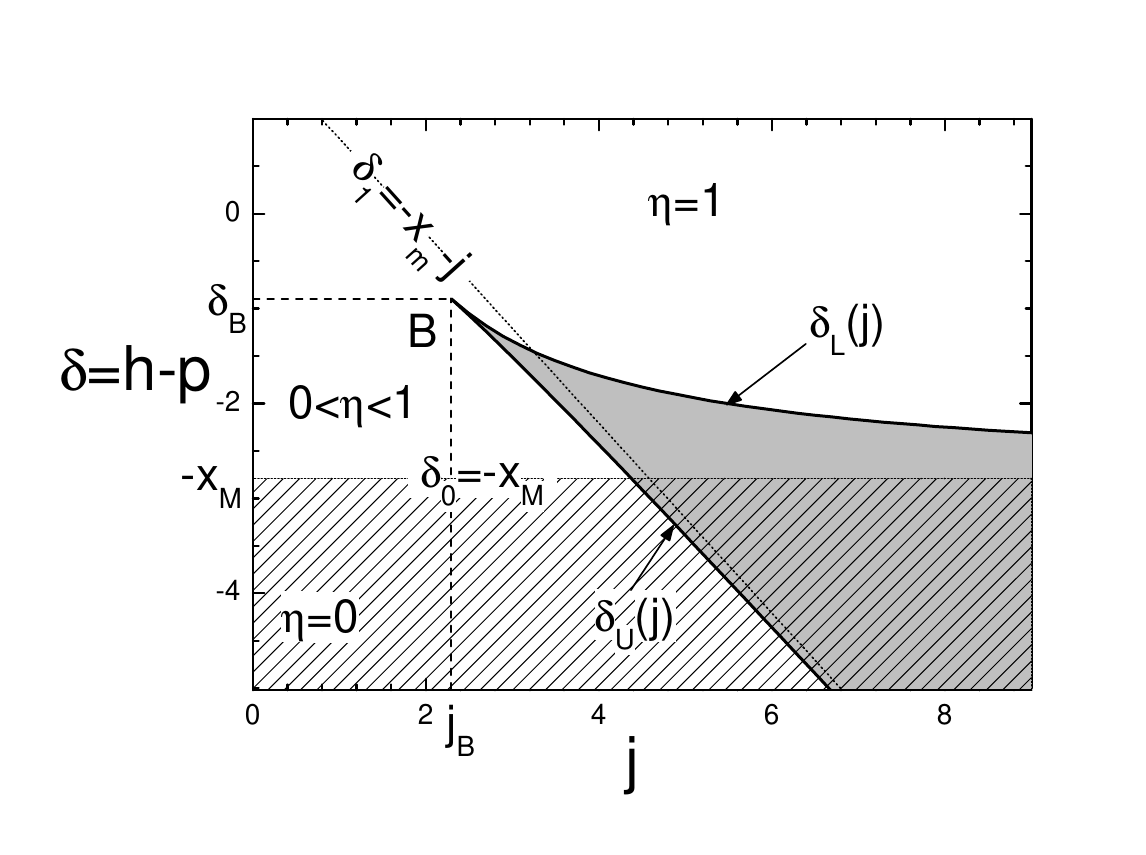}
\caption[]{{\em Triangular pdf of unitary variance and a maximum at $x_B=-1$: 
customers phase diagram. 
}}
\label{fig:triangM-1PhDiagrams}
\end{figure}

Summarizing, the customers phase diagram for pdfs with compact supports have two supplementary lines with respect to that with unbounded supports. They indicate the boundary of the {\em viability region} (no market exists below this line) and the {\em saturation} boundary (above which all the customers are buyers). Figure \ref{fig:triangM-1PhDiagrams} presents the customers' phase diagram for our example corresponding to the triangular pdf of unitary variance (\ref{eq:triang_distribution}), with a maximum at $x_B=-1$.

\subsection{Pdfs with fat tails}
\label{sec:fat}
Fat-tail distributions are characterized by the fact that the pdf $f$ has a slow decrease at large values of $x$,
so that the variance is infinite - or even the mean is infinite. 
Equivalently this occurs if
$\Gamma(\eta)$ diverges 'too fast' to $-\infty$ when $\eta \rightarrow 0$.
In the case of the logistic, $\Gamma \sim \log \eta$; for the Gaussian, $\Gamma \sim - \sqrt{-2 \log \eta}$;
for a power law, $\Gamma \sim - \frac{1} {\eta^b}$.
This suggest to consider the general following smooth behavior:
\begin{equation}
\mbox{as $\eta \rightarrow 0$: } \; \Gamma(\eta) \sim - K (-\log \eta)^a \frac{1} {\eta^b},
\label{eq:eta->0}
\end{equation}
with the constant $K>0$, $a \geq 0$ and $b \geq 0$ ($ab \neq 0$). 
The fat-tail case corresponds then to
 $a=0$ and $b \geq 1$ in the above equation (\ref{eq:eta->0}). 

A particular example of a fat tail distribution is a pdf with a power law decrease, which for large $x$ behaves like:
\begin{equation}
f(x) \sim x^{-(1+\mu)}
\label{eq:powerlaw}
\end{equation}
with $\mu \geq 0$. Then, for small $\eta$, $\Gamma \sim - \frac{1} {\eta^b}$ with $b=1/\mu$, 
so that $b\geq 1$ means $\mu \leq  1$. 
For $\mu <1$, not only the variance but also the mean value of the random variable $x$ is infinite.

For fat-tails distributions one has to look at finite size effects: it is no more possible to take directly
the large $N$ limit and make use of the central limit theorem:
quantities like $(1/N) \sum_i G(x_i)$ for any function $G$ will be dominated
by rare events, that is by the largest values encountered in the population of (large but finite) size $N$.
There is, however, no difficulty in doing this analysis: the results are obtained by doing {\em as if}
the pdf had a finite support, the upper bound $x_M$ being given as an increasing function of $N$
(for an introduction to statistics with fat tails, see e.g. \cite{BouchaudPotters}).

Since we considered compact supports in the preceding section, 
we concentrate here on the marginal case $\mu=b=1$, which can be analyzed as a limiting case
of distributions with infinite support.

\begin{figure}
\centering
\includegraphics[width=0.7\textwidth]{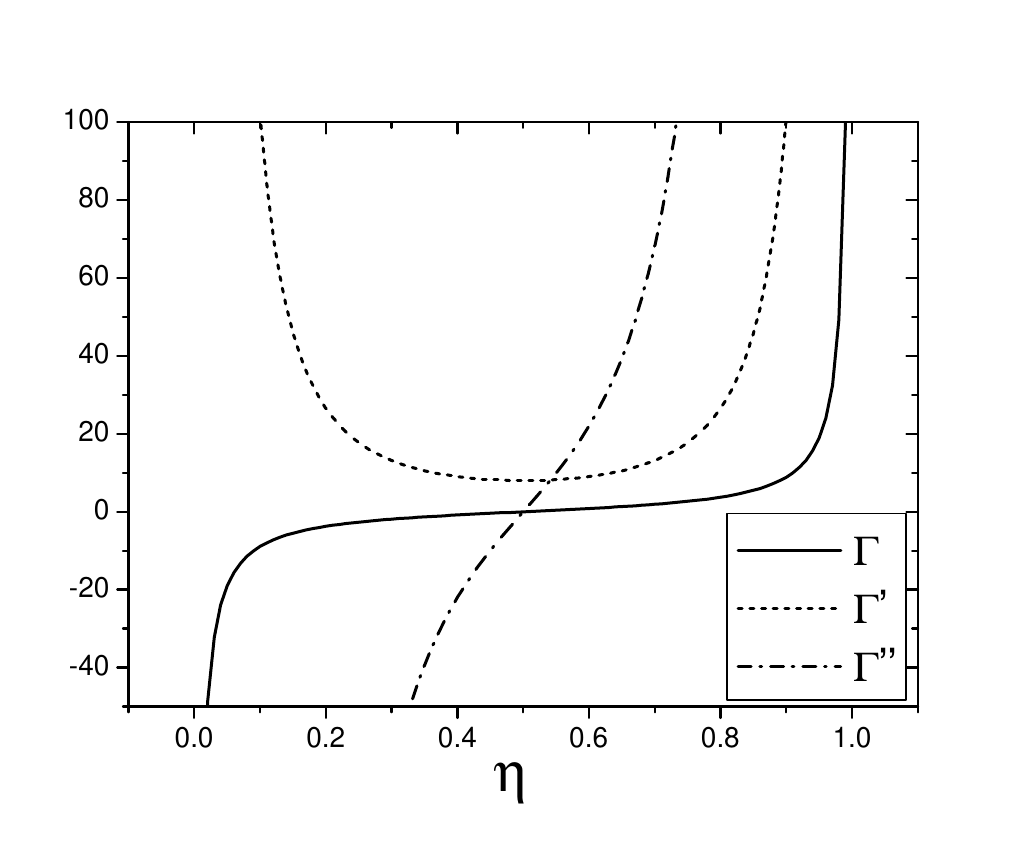}
\caption[]{{\em $\Gamma$ function and derivatives corresponding to the function (\protect{\ref{eq:Gamma_b=1}}).}}
\label{fig:FatTailGammas}
\end{figure}

\begin{figure}
\centering
\includegraphics[width=0.7\textwidth]{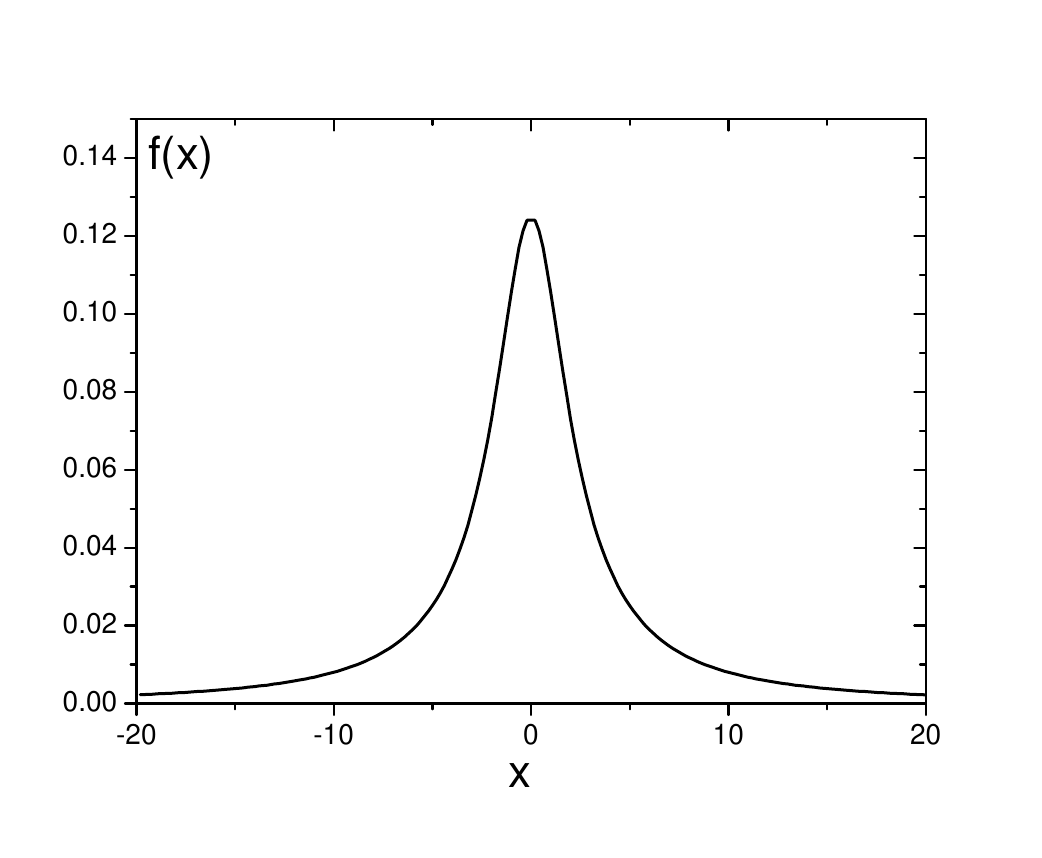}
\caption[]{{\em Pdf corresponding to equation (\protect{\ref{eq:pdf_b=1}}).}}
\label{fig:FatTailPdf}
\end{figure}

For $\mu=1$, $f(x)$ does not have a finite variance. Then, the value of $\sigma$ that defines the normalized variables (\ref{eq:reduced_var}) may be any (finite) measure of the width of the distribution, as for example, the value of $x$ at which $f(x)$ is equal to half its maximum. Let us discuss this marginal case on a simple example (see figure \ref{fig:FatTailGammas}):
\begin{equation}
\Gamma \equiv - \frac{1}{\eta} + \frac{1}{1-\eta},
\label{eq:Gamma_b=1}
\end{equation}
corresponding to the cumulative function:
\begin{equation}
F(z) = \frac{1}{2} - \frac{1}{z} + \mbox{sgn}(z) \sqrt{\frac{1}{z^2} + \frac{1}{4}}.
\label{eq:F_b=1}
\end{equation}
The corresponding pdf is,
\begin{equation}
f(x) = \frac{1}{x^2} [1 - \frac{2}{\sqrt{4+x^2}}],
\label{eq:pdf_b=1}
\end{equation}
as represented on figure \ref{fig:FatTailPdf}. Since this is a symmetric distribution, $\eta_B=1/2$, and one finds (see equations (\ref{eq:etaB}), (\ref{eq:jB}) and (\ref{eq:deltaB}) ) that the point $B$ in the customers phase diagram is $j_B=\Gamma'(\eta_B)=8$, $\delta_B=-4$. Notice that, like for any monomodal distribution (satisfying thus hypothesis H1), $\Gamma(\eta)$ is convex for $\eta>\eta_B$ and concave for $\eta < \eta_B$, with as before $\eta_B$ being the inflexion point. The supply function has thus the generic behavior described in Section \ref{sec:demand} even for fat-tail distributions. 

\subsection{Aggregate demand for multimodal pdfs}  
\label{sec:app_multimodal}

\subsubsection{Smooth pdfs: generic properties}
\label{sec:app_multimodal_smooth}

In this section we consider the behavior of the application $\delta \rightarrow \eta^d(\delta)$ 
in the case of a smooth multimodal pdf with support on $]-\infty, +\infty[$.  
The discussion Section~\ref{sec:demand_msr}, based on convexity arguments, can be extended to describe the phase diagram
for the aggregate demand in the multimodal case.

The minimal hypotheses we consider are the following.
\begin{itemize}
\item HA0.
The pdf $f(x)$ is continuous with a finite number $K \geq 2$ of $x$-values, 
$-\infty < x_B^K < x_B^{K-1} < ... < x_B^1 < \infty$,
for which $f$ has a (possibly local) bounded maximum,
\begin{equation}
f(x_B^k) \; < \infty \;\;\;\; \; k=1,...,K.
\label{eq:maxbounded}
\end{equation}
For simplicity we assume also that the pdf is not constant 
on any interval of finite length. Actually, the discussion 
can be easily extended to less
regular pdfs (in particular piecewise continuous pdfs), 
and pdfs constant on some intervals,
but to keep the discussion shorter will leave that to the interested reader
(in the case of a monomodal pdf, see the discussion on compact supports,
and for the bimodal case see also below, Section~\ref{sec:app_2diracs}, the singular case of a distribution composed of two
Diracs).
\item HA1.
When considering smoother functions, we will assume $f$ to be twice continuously differentiable,
so that in particular it has a zero derivative at every maximum and every minimum.
\end{itemize}

Let us denote by $x_C^k$ the location of the minimum between
$x_B^{k+1}$ and $x_B^k$. We assume $f > 0$ everywhere on its support except possibly at some minima, and
$f(x)$ goes to zero as $x \rightarrow \pm \infty$.
By convention we set $x_C^0 = +\infty$ and $x_C^K = -\infty$ (and we may write $f(x_C^0)=f(x_C^K)=0$).

In the monomodal case, we have seen that the critical value of $j$ for the appearance
of several solutions is $j_B = 1/f(x_B)$. Here we will see that the relevant critical values are
\begin{equation}
j_B^k \equiv \frac{ 1 }{ f(x_B^k) } \;\;\;\; \; k=1,...,K
\label{eq:jb}
\end{equation}
and also
\begin{equation}
j_C^k \equiv \frac{ 1 }{ f(x_C^k) } \;\;\;\; \; k=1,...,K-1
\label{eq:jc}
\end{equation}
(and it will be convenient to define as well $j_C^0= j_C^K = \infty$).

Consider now the inverse demand at a given value of $j$.
The ensemble of equilibria $\delta^d(\eta)$ for $\eta \in [0,1]$
is the subset of the ensemble of solutions of~(\ref{eq:h-p}) for which $\delta$ increases ($p$ decreases)
as $\eta$ increases. As for the monomodal case we study the function of $\eta$ defined by~(\ref{eq:h-p})
for any given $j$,
$\delta(\eta)={\cal D}(j;\eta)$. By continuity of the function $\Gamma$,
$\delta (\eta)$ is a continuous function of $\eta \in [0,1]$. As $\eta \rightarrow 0$,
$\delta \rightarrow - \infty$, and as $\eta \rightarrow 1$, $\delta \rightarrow + \infty$.
Increasing $\eta$ from $0$, $\delta(\eta)$ increases. Similarly, decreasing $\eta$ from $\eta=1$,
$\delta(\eta)$ decreases. Since $f(x)$ is continuous, $\Gamma(\eta)$ is continuously differentiable,
with $\Gamma'(\eta) \equiv d\Gamma(\eta) / d\eta = 1/f(x)$ at $x= - \Gamma(\eta)$. Hence $\Gamma'(\eta)$
has (local) minima at values of $\eta$ given by
\begin{equation}
\Gamma'(\eta_B^k) = \frac{ 1 }{ f(x_B^k) } \;\;\;\; \; k=1,...,K
\label{eq:etabk}
\end{equation}
and (local) maxima  at values of $\eta$ given by
\begin{equation}
\Gamma'(\eta_C^k) = \frac{ 1 }{ f(x_C^k) } \;\;\;\; \; k=1,...,K-1
\label{eq:etack}
\end{equation}
For a smooth enough pdf, the $\eta_B^k$ and $\eta_C^k$s are inflexion points for $\Gamma$.
Note that for any $k=1,...,K$,  $\eta_C^{k-1} < \eta_B^k < \eta_C^k$.

The most important remark is that
$\Gamma(\eta)$ is strictly concave on every interval $]\eta_C^{k-1},\; \eta_B^k[,\; k=1,...,K$, and 
strictly convex on every interval $]\eta_B^k,\; \eta_C^k[,\; k=1,...,K$.
Then as $\eta$ varies on $[\eta_C^{k-1},\; \eta_B^k]$, the function ${\cal D}(j,\eta) = \Gamma(\eta) - j \eta$ 
has, at some value $\eta_L^k(j)$, a maximum $\delta_L^k(j)$ which is by definition the Legendre transform of
$\Gamma(\eta)$ restricted to $[\eta_C^{k-1},\; \eta_B^k]$. 
Similarly, on $[\eta_B^k,\; \eta_C^k]$,  ${\cal D}(j,\eta)$ has, at some value $\eta_U^k(j)$,  a minimum $\delta_U^k(j)$,
the Legendre transform of $\Gamma(\eta)$ restricted to $[\eta_B^k,\; \eta_C^k]$.

Depending on the value of $j$ compared to the values $j_B^k, j_C^k$, these min and max may be reached either at
a boundary of an interval, or in the interior. More precisely:
\begin{eqnarray}
j < j_B^k, & \eta_L^k = \eta_U^k = \eta_B^k \\
j_B^k < j < j_C^{k}, & \eta_B^k < \eta_U^k < \eta_C^k \\
j_B^k < j < j_C^{k-1}, & \eta_C^{k-1} < \eta_L^k < \eta_B^k \\
j_C^{k} < j , & \eta_L^{k+1} = \eta_U^{k} = \eta_C^{k} 
\label{eq:app_julbc}
\end{eqnarray}
(and $\eta_U^k$ increases from $\eta_B^k$ to $\eta_C^k$ as $j$ increases from
$j_B^k$ to $j_C^k$, whereas $\eta_L^k$ decreases from $\eta_B^k$ to $\eta_C^{k-1}$ as $j$ increases from
$j_B^k$ to $j_C^{k-1}$).
In the case of a continuously differentiable pdf, 
every Legendre transform $\eta_{\Lambda}^k(j), \; \Lambda=L,U$
satisfies the marginal stability equation,
\begin{equation}
\frac{\partial {\cal D}(j, \eta)}{\partial \eta}|_{\eta=\eta_{\Lambda}^k(j)}= 0.
\label{eq:app_margin}
\end{equation}
One should note that $\eta_{U,L}^k$ and $\delta_{U,L}^k$ depend on $j$ 
(and on the function $\Gamma(.)$), but not on $h$ or $p$. 

Now for $j< j_B \equiv \min_k j_B^k$, every min and max are reached at the corresponding value
$\eta_B^k$: this means that
there is no intermediate regime with a decreasing behavior of $\delta(\eta)$ as $\eta$ increases, 
hence $\delta^d(\eta)={\cal D}(j,\eta)$, 
uniquely defined, is a continuously increasing function
of $\eta \in [0,1]$. For $j> j_B$, there is at least one $k$
where the maximum $\delta_L^k(j)$ is reached for 
$\eta=\eta_L^k(j) < \eta_B^k$,
and the minimum $\delta_U^k(j)$ is reached for $\eta=\eta_U^k(j) > \eta_B^k$, so that
there is at least one finite interval of $\eta$ on which the function
${\cal D}(j,\eta)$
decreases with $\eta$, and thus does not correspond to an
economic equilibrium. Hence the demand $\eta^d(\delta)$ has at least two branches. 

In the plane $(j, \delta)$, the boundaries of the multiple solutions regions 
are thus given by the fonctions $\delta_{\Lambda}^{k}(j) =  {\cal D}(j,\eta_{\Lambda}^{k}(j)),\;\; \Lambda= L,U$, 
which are the graphs of all the 
branches of the Legendre transform of $\Gamma$. 
By construction of the Legendre transform, every branch $\delta=\delta_U^k(j)$ is a concave
curve, and every branch $\delta=\delta_L^k(j)$ is a convex curve, and, under the smoothness hypothesis HA1, along each branch $\Lambda=L,U$, 
\begin{equation}
\frac{ d \delta_{\Lambda}^{k}(j) }{dj } = \frac{d  {\cal D}(j,\eta_{\Lambda}^k(j))}{d j}=-\eta_{\Lambda}^k(j).
\label{eq:universal_slope}
\end{equation}
Recall that $\eta_{\Lambda}^k$ is the value of $\eta$ for the solution
which is marginally stable on this boundary.

\begin{figure}
\centering
{\rotatebox{270}{
\includegraphics[width=0.8\textwidth,angle=90]{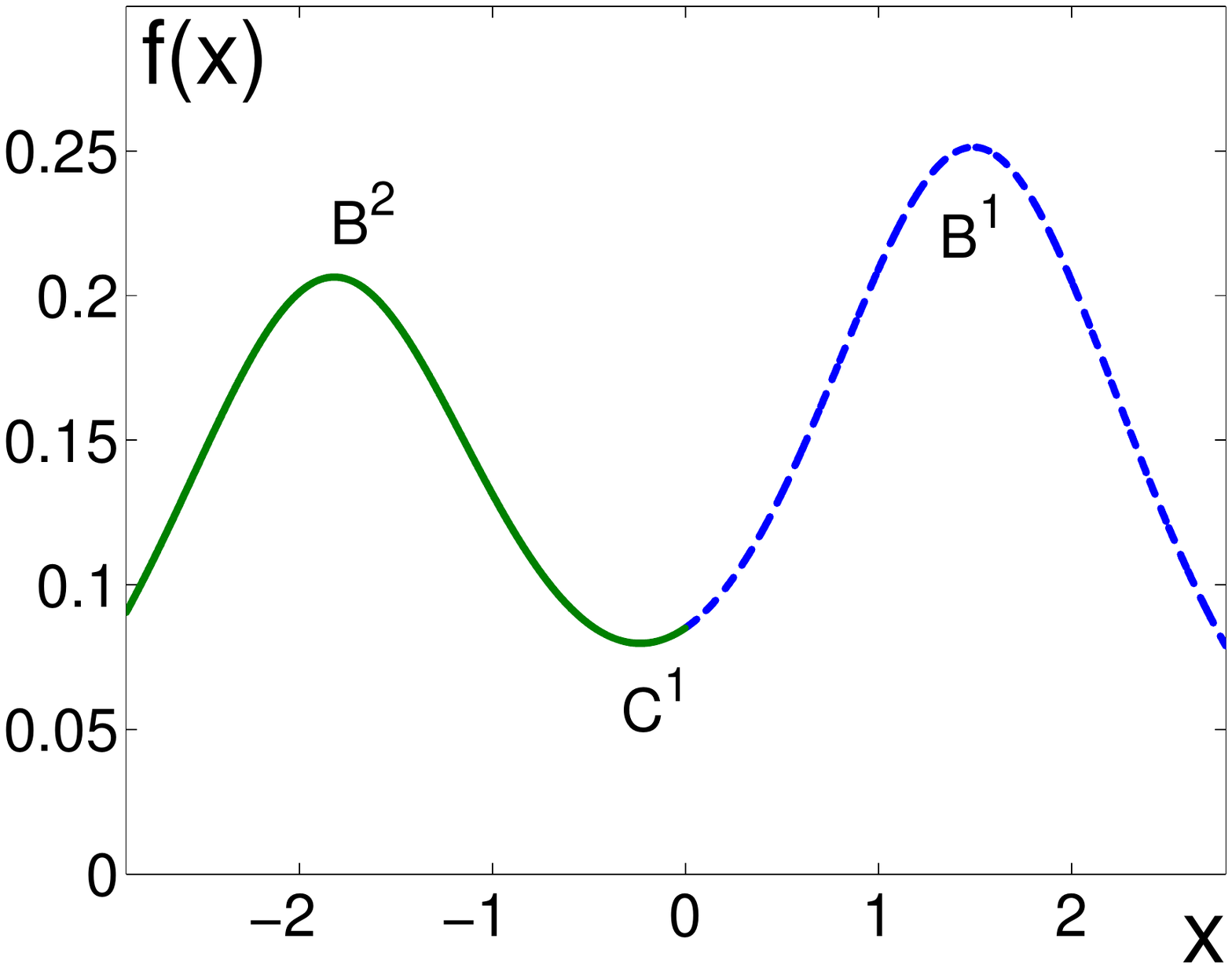}
}}
\caption[]{\em An example of bimodal pdf. }
\label{fig:bimodal_pdf}
\end{figure}

\begin{figure}
\centering
{\rotatebox{270}{
\includegraphics[width=0.8\textwidth,angle=90]{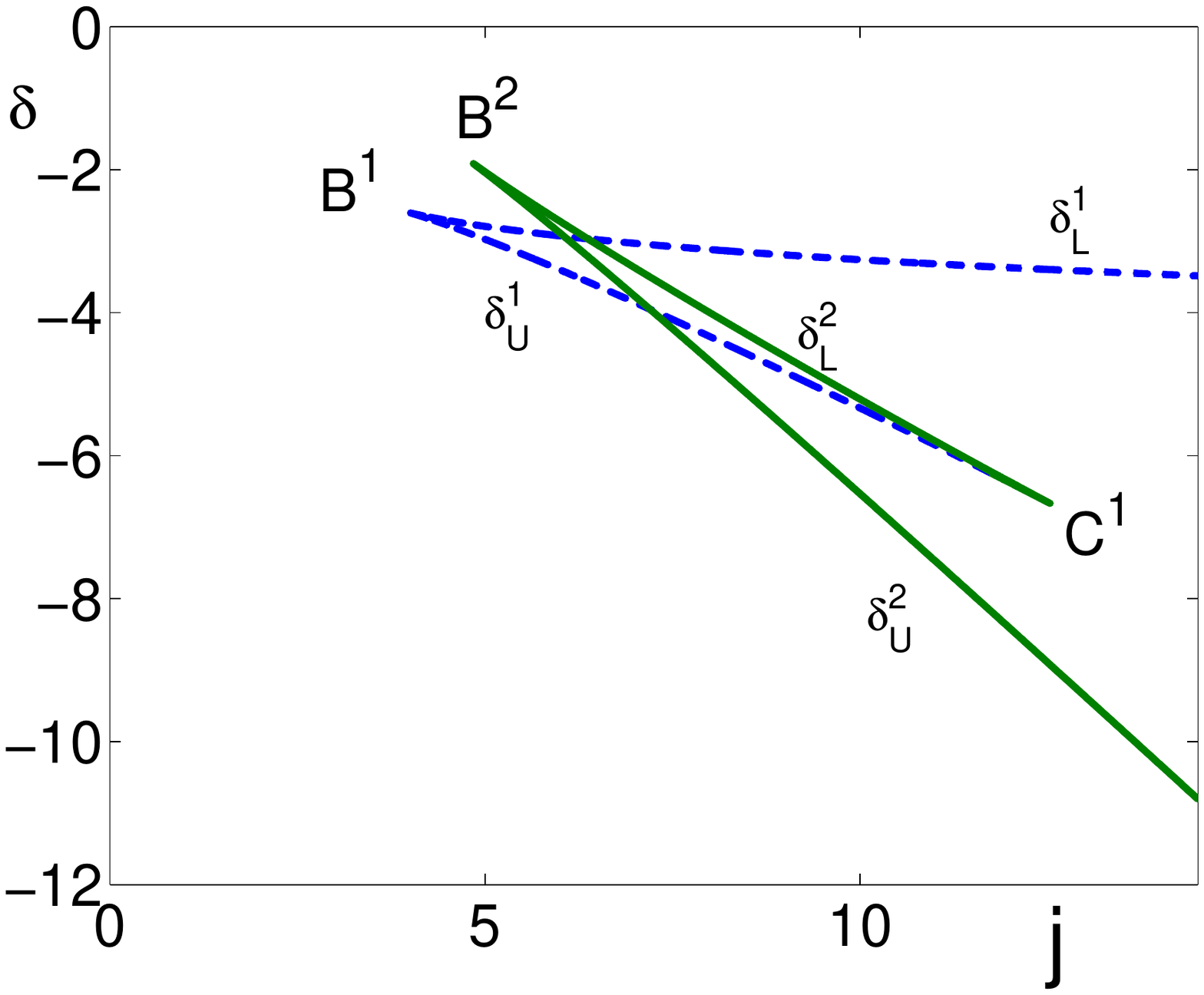}
}}
\caption[]{\em Phase diagram (aggregate demand) for the case of the smooth bimodal pdf shown on figure~\ref{fig:bimodal_pdf}.}
\label{fig:bimodal}
\end{figure}

These boundaries can be easily drawn for any distribution making use of a parameterization by $s$
(or equivalently $x \equiv -s$): from the basic equations $\eta= 1 - F(-s)$ where $F$ is the cumulative of 
the pdf $f$, $s = \Gamma(\eta)$, and $\Gamma'(\eta)=1/f(-s)$; with the marginal stability
condition~(\ref{eq:app_margin}) which gives $j=\Gamma'(\eta)$, the locus of marginal stability
is then given in the plane $(j,\delta)$ by the parameterized curve
\begin{eqnarray}
{\rm for}\; x \in {\rm support}( f ), & &  \nonumber \\
j & =&  1/f(x) \\
\delta & =  & -x - \frac{1 - F(x)}{f(x)} 
\label{eq:app_margin_curve}
\end{eqnarray}
This is this representation that we have used to draw the phase diagram, figure \ref{fig:bimodal}, for the particular example 
of the bimodal distribution shown on figure~\ref{fig:bimodal_pdf}.


The domain of multiple solutions can then be described as follows.
The phase diagram is a kind of superposition of
diagrams associated to mono-modal phase diagrams,
every maximum (every 'bump' in the pdf) $k$ being responsible of the 
appearance of a domain of multistability: 
when $j$ becomes larger than $j_B^k$, a continuous solution split into two solutions,
with 
a lower solution $\eta^d(j,\delta) \leq \eta_L^k(j) < \eta_B^k$
and $\delta \leq \delta_L^k$,
and an upper one with $\eta^d(j,\delta) \geq \eta_U^k(j) > \eta_B^k$ and $\delta \geq \delta_U^k$
(see figure \ref{fig:bimodal}). 
When $j$ becomes larger than $j_C^k$, this bump is no more 'seen'.
Since a minimum of the pdf, if not at a boundary, is in between two maxima, such
an intermediate solution may exist either because of one bump or the other - or both.

The branch $\delta = \delta_U^k(j)$ has thus as left end point,
$B^k \equiv (j_B^k, \delta_B^k= {\cal D}(j_B^k,\eta_B^{k}) )$, and as right endpoint (if $j_C^k$ is finite),
$C^k\equiv (j_C^k, \delta_C^k= {\cal D}(j_C^k,\eta_C^{k}) )$. 
$B^k$ is the merging point of $\delta_U^k$ and $\delta_L^k$, and $C_k$ the merging point of
$\delta_U^k$ and $\delta_L^{k+1}$. Since $\delta_L^k$ and $\delta_L^{k+1}$ must be both above
$\delta_U^k$, these two branches must intersect one another for some value of $j=j_{BC}^k$ between
$j_B^k$ and $j_C^k$: there is thus coexistence of three solutions
in the triangular-like domain bounded below by $\delta_U^k$
(or $\max(\delta_U^k, \;\delta_U^{k+1})$ if $B^k$ is below the branch $\delta_U^{k+1}$), and above by
$\delta_L^k$ for $j \leq j_{BC}^k$, and by $\delta_L^{k+1}$ for $j \geq j_{BC}^k$.

In the smooth case (HA1), at every bifurcation point $B^k$, resp. $C^k$ where two boundaries merge, according
to~\ref{eq:universal_slope} 
there is a common slope $-\eta_B^k$, resp. $-\eta_C^k$ . 

\begin{figure}
\centering
{\rotatebox{270}{
\includegraphics[width=0.8\textwidth,angle=90]{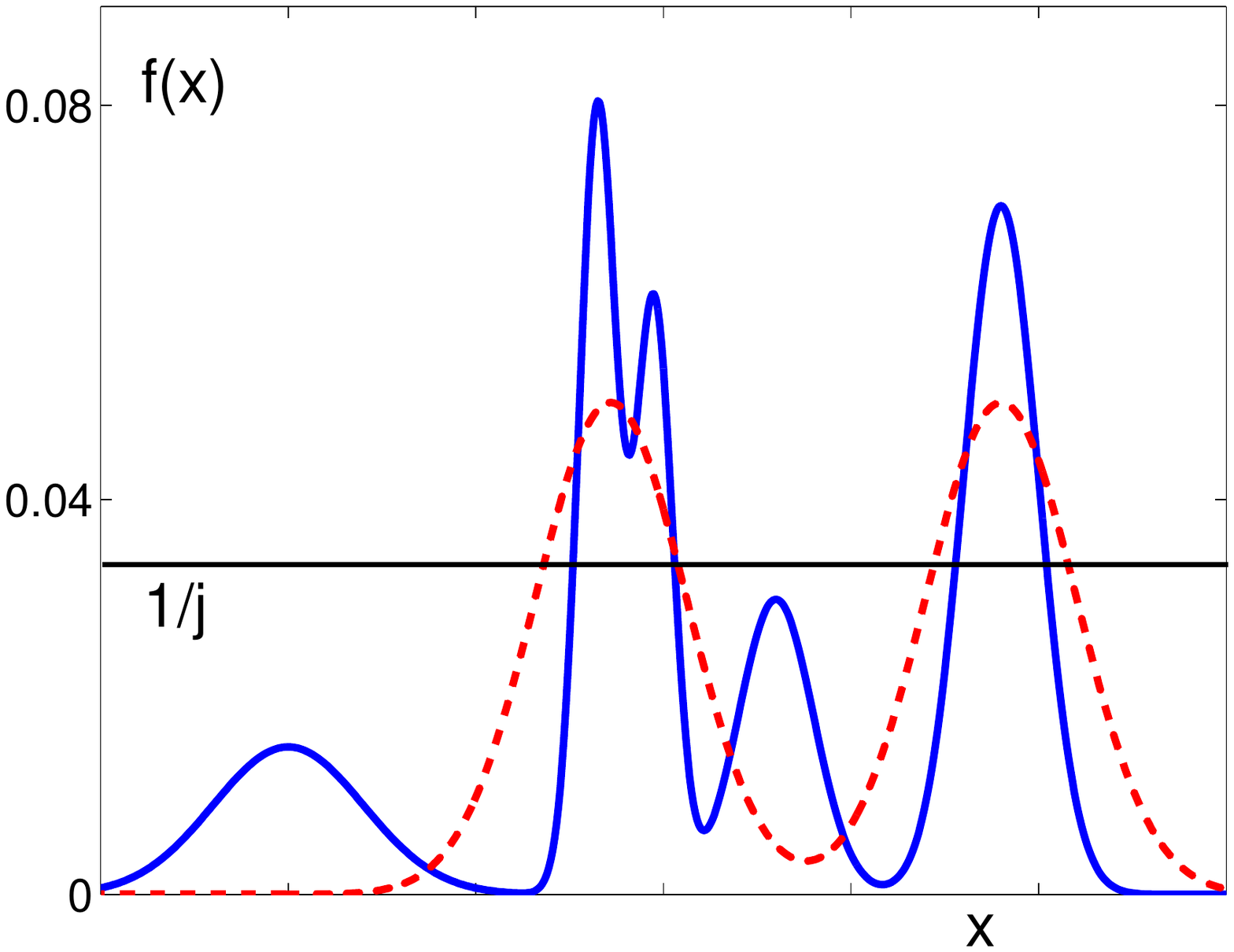}
}}
\caption[]{\em Examples of multimodal pdfs. At a given value of $j=J/\sigma$, the qualitative
properties are obtained by looking at the intersection of the horizontal line $y= 1/j$ with the 
graph of the pdf, $y=f(x)$: for the particular value of $j$ corresponding to the
horizontal line on this figure,  
the two pdfs lead to the same qualitative properties of the Demand.}
\label{fig:app_multimodal}
\end{figure}

One may say that the pdf is probed at different scales for different values of $j$.
Consider the graph $y=f(x)$. Every maximum below the line
$y=1/j$ is not seen (it does not change the structure of the solution), whereas a set of maxima higher than $1/j$, 
but joined by minima 
where $f$ is still higher than $1/j$, is seen as a single global bump.
This gives in particular that for $j > j_B$, the number of solutions is equal to one 
plus the number of times the line $y=1/j$
cut the graph $y=f(x)$ at points where $f$ is increasing. 
Note that this does not give the number of solutions for a given value of $\delta$.
On figure~\ref{fig:app_multimodal}, two pdfs are shown; the intersection of the
graph $y=f(x)$ with the line $y=1/j$ gives the structure of the demand at this particular value of
$j$ (in the case illustrated on the figure, the demand has $3$ solutions for the two pdfs).

\subsubsection{A degenerate case: 2 Dirac}
\label{sec:app_2diracs}

\begin{figure}
\centering
{\rotatebox{270}{
\includegraphics[width=0.8\textwidth,angle=90]{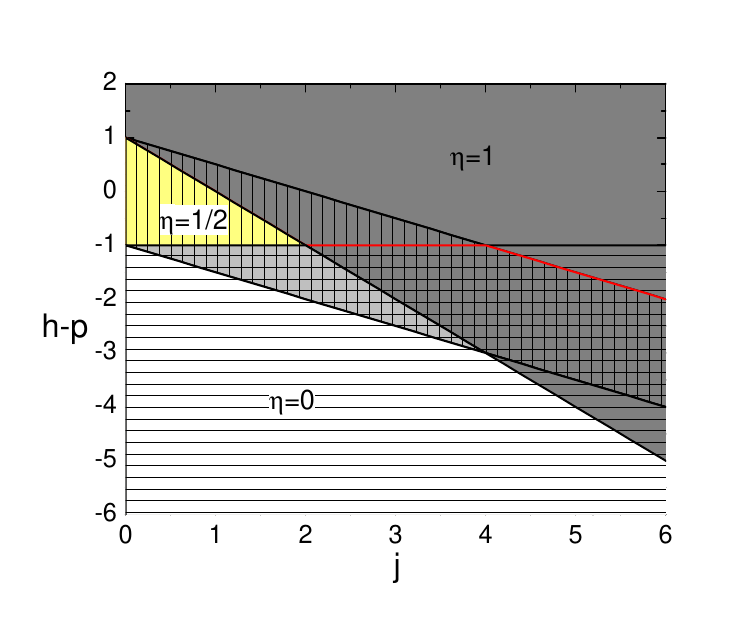}
}}
\caption[]{\em Phase diagram (aggregate demand) for the case of a bimodal pdf composed of two Dirac peaks.}
\label{fig:2diracs}
\end{figure}

Let us consider the particular case of an IWP distribution given by two
Delta pics: $x_i= \pm x_0$ with equal probability ($x_0 = 1/\sqrt{2}$ since the variance of $f$
is normalized to $1$). For $j=0$,
one has clearly $\eta=0, 1/2$ or $1$ depending on $\delta < - x_0$, $-x_0 < \delta < + x_0$
or $\delta > x_0$. For $j> 0$, obviously $\eta$ can still take only these three
values. One gets easily the domain of existence and stability of these solutions,
$\eta=0, 1/2, 1$, by direct inspection of the equation~(\ref{eq:fp}).
The resulting phase diagram is shown on figure~\ref{fig:2diracs}. 

This phase diagram for a singular distribution can also be understood 
by comparison with the predicted phase diagram for a continuous distribution.
In the present case, the two maxima have equal height, $+\infty$, which gives $j_B^1=j_B^2=0$,
in agreement with the fact that boundary lines meet at $j=0$. The minimum
between the two maxima is at $f=0$, hence $j_C=\infty$: the domain of stability of
the intermediate solution $\eta=1/2$ extends to infinity, as it is the case whenever a minimum is 
at $f(x_C^1)=0$. The marginal stability lines are straight lines - hence, marginally concave and convex curves -,
with slopes $0, 1/2$ and $1$ corresponding to the values of the solution
marginally stable on the boundary, in agreement with~(\ref{eq:universal_slope}).
Since here there is no continuity in the demand at the singular points $B^1=(0,-1), B^2=(0,1)$,
two branches do not merge with a common slope: besides the fact that the demand can take only three values,
this is the only place where the non smoothness of the pdf gives a feature of the phase diagram
qualitatively different from what is obtained for a smooth pdf.

\end{document}